\documentclass{ieeeaccess}
\usepackage{cite}
\usepackage{amsmath,amssymb,amsfonts}
\usepackage{algorithmic}
\usepackage{algorithm}
\usepackage{graphicx}
\usepackage{textcomp}
\usepackage{booktabs}
\usepackage{interval}
\usepackage{etoolbox}

\DeclareMathOperator*{\argmax}{arg\,max}
\DeclareMathOperator*{\argmin}{arg\,min}

\def\BibTeX{{\rm B\kern-.05em{\sc i\kern-.025em b}\kern-.08em
    T\kern-.1667em\lower.7ex\hbox{E}\kern-.125emX}}
\begin{document}
\history{Date of publication xxxx 00, 0000, date of current version xxxx 00, 0000.}
\doi{XX.XXXX/XXXXXX.XXXX.DOI}

\title{hARMS: A Hardware Acceleration Architecture for Real-Time Event-Based Optical Flow}

\author{\uppercase{Daniel C. Stumpp}\authorrefmark{1},
\uppercase{Himanshu Akolkar}\authorrefmark{2}, \uppercase{Alan D. George}\authorrefmark{1}, \uppercase{\\and Ryad B. Benosman}\authorrefmark{2,}\authorrefmark{3,}\authorrefmark{4}}

\address[1]{University of Pittsburgh, Pittsburgh, PA 15213, USA}
\address[2]{Biomedical Science Tower 3, University of Pittsburgh, Pittsburgh, PA 15260, USA}
\address[3]{CNRS, UMRS 968, UMR 7210, INSERM UMRI S 968, Institut de la Vision, Sorbonne Université (UPMC University Paris 06), 75012 Paris, France}
\address[4]{Robotics Institute, Carnegie Mellon University, Pittsburgh, PA 15213, USA}

\tfootnote{This research was supported by the NSF Center for Space, High-performance, and Resilient Computing (SHREC) industry and agency members and by the IUCRC Program of the National Science Foundation under Grant No. CNS-1738783.}

\markboth
{D. Stumpp \headeretal: hARMS: A Hardware Acceleration Architecture for Real-Time Event-Based Optical Flow}
{D. Stumpp \headeretal: hARMS: A Hardware Acceleration Architecture for Real-Time Event-Based Optical Flow}

\corresp{Corresponding authors: Daniel C. Stumpp (dcs98@pitt.edu) and Ryad B. Benosman (benosman@pitt.edu)}

\begin{abstract}
Event-based vision sensors produce asynchronous event streams with high temporal resolution based on changes in the visual scene. The properties of these sensors allow for accurate and fast calculation of optical flow as events are generated. Existing solutions for calculating optical flow from event data either fail to capture the true direction of motion due to the aperture problem, do not use the high temporal resolution of the sensor, or are too computationally expensive to be run in real time on embedded platforms. In this research, we first present a faster version of our previous algorithm, ARMS (Aperture Robust Multi-Scale flow). The new optimized software version (fARMS) significantly improves throughput on a traditional CPU. Further, we present hARMS, a hardware realization of the fARMS algorithm allowing for real-time computation of true flow on low-power, embedded platforms. The proposed hARMS architecture targets hybrid system-on-chip devices and was designed to maximize configurability and throughput. The hardware architecture and fARMS algorithm were developed with asynchronous neuromorphic processing in mind, abandoning the common use of an event frame and instead operating using only a small history of relevant events, allowing latency to scale independently of the sensor resolution. This change in processing paradigm improved the estimation of flow directions by up to 73\% compared to the existing method and yielded a demonstrated hARMS throughput of up to 1.21 Mevent/s on the benchmark configuration selected. This throughput enables real-time performance and makes it the fastest known realization of aperture-robust, event-based optical flow to date.
\end{abstract}


\titlepgskip=-15pt

\maketitle

\section{Introduction}
\label{sec:introduction}

The emergence of event-based vision sensors has led to the development of new applications and algorithms that are able to leverage the high temporal resolution that these sensors provide. One such application is the computation of optical flow. Traditional optical flow algorithms such as the Horn and Schunk \cite{Horn1981DeterminingFlow} and Lucas-Kanade \cite{lucas1981iterative} methods operate on traditional camera frames and thus are not suitable to take advantage of the high temporal resolution and asynchronous characteristics of event-based vision sensors. Many new methods for computing optical flow from event-based sensors have been developed to capitalize on the unique characteristics of these sensors. One such method is a modified version of the Lucas-Kanade method that uses the events to determine spatial and temporal gradients \cite{Benosman2012AsynchronousFlow}. Another method, uses the derivative of the regularized surface of events to estimate the magnitude and direction of an object's motion \cite{Benosman2014Event-basedFlow}. Both methods discussed are susceptible to the aperture problem of optical flow, i.e, they will produce flow vectors that are normal to the moving edge regardless of the true direction of motion. This is due to the fact that both methods view only a local region of events and do not consider all the events produced by an object moving through the scene. 

There are some methods of event-based optical flow calculation that have been able to address the aperture problem and compute the true optical flow using event data. One such method, known as EV-FlowNet, is presented in \cite{Zhu2018EV-FlowNet:Cameras}. This method uses a self-supervised neural network to generate the optical flow using frame based accumulation of events. This method accumulates events over a certain time window before calculating the flow for the whole frame. It is also dependent on the use of quality grayscale images generated by the DAVIS event camera for training, making it less adaptable to varied and unpredictable visual scenes \cite{Zhu2018EV-FlowNet:Cameras}. Akolkar et. al.~\cite{Akolkar2020Real-timeFlow} proposed an unsupervised event per event spatial pooling of local-flow computations to solve the aperture problem while calculating flow asynchronously using only generated temporal contrast events. They present a method referred to as aperture robust multi-scale (ARMS) for computing optical flow from an event stream \cite{Akolkar2020Real-timeFlow}. Details of the ARMS algorithm are discussed further in Section~\ref{sec:multiscale-pooling}.

In this research, we propose a redesigned and optimized ARMS algorithm referred to as faster ARMS (fARMS), and analyze time complexity when compared to the original ARMS algorithm. We then present a hardware acceleration architecture of the fARMS flow algorithm using a hybrid system-on-chip (SoC) embedded platform containing a field programmable gate array (FPGA). We will refer to this architecture as hardware ARMS (hARMS). fARMS and hARMS were developed to complement the asynchronous nature of event-based vision sensors and the events they output. The hARMS architecture allows for flexible configuration based on application specific needs and provides significant improvements in latency and throughput compared to both the existing ARMS flow algorithm and the fARMS software baseline. This improved performance allows for real-time operation in a variety of visual scenarios, opening up the possibility of more widespread optical-flow-based application deployment on embedded platforms.

Although there are a variety of event-based optical flow techniques, relatively little research has been done on hardware acceleration of them. This is due in part to the fact that some of the local-flow algorithms are able to perform near real time without hardware acceleration because they only consider a small amount of local data. There has, however, been some related research in this area.

A block-matching optical flow algorithm for event-based sensors was implemented on an FPGA by Liu and Delbruck \cite{Liu2017Block-matchingImplementation}. This implementation, however, was found to perform poorly in real-world scenes and was therefore expanded on in \cite{Liu2019AdaptiveSensors} to improve performance. It is estimated in \cite{Liu2019AdaptiveSensors} that the improved design would require 100k look-up tables (LUTs) and 35k flip-flops (FFs) when implemented on an FPGA using a similar architecture as proposed in \cite{Liu2017Block-matchingImplementation}. Block matching was shown, in some cases, to find the true optical flow, however this was dependent on a predetermined block size parameter and the dynamics of the scene \cite{Liu2019AdaptiveSensors}. Although this method can overcome the aperture problem in some cases, it still fails to fully make use of the high temporal resolution of event-based cameras. The algorithm is not asynchronous, but rather operates on time slices of accumulated events, therefore sacrificing temporal resolution. 

Another example of event-based optical flow acceleration using an FPGA is presented in \cite{Aung2018Event-basedFPGA}. This research presents an FPGA implementation for a modified version of the iterative derivative of the surface of events algorithm presented in \cite{Benosman2014Event-basedFlow}. The algorithm derives the surface of events after a temporal regularization to asynchronously estimate the flow, and is capable of performing at a throughput of 2.75 Mevt/s. However, the overall throughput of the system is limited to 1.46 Mevt/s due to the pre-processing stage \cite{Aung2018Event-basedFPGA}. The design was implemented using a Xilinx Spartan 6 LX150 FPGA on an Opal Kelly XEM6010 board and required 3794 logic slices, 138 block RAMs (BRAM), and 16 DSPs when using a 304$\times$240~$px$ resolution ATIS sensor.

While results presented in \cite{Aung2018Event-basedFPGA} show impressive throughput, allowing for real-time, asynchronous operation, it makes no attempt to address the aperture problem. The design presented also has high BRAM requirements because recent events at each pixel location are stored. This indicates that the architecture would scale poorly as the resolution of the event-based vision sensor increases, drastically increasing BRAM requirements and potentially exceeding the available resources on many embedded FPGA platforms.  Finally, an implementation of the surface of events approach \cite{Benosman2014Event-basedFlow} using spiking neural networks and a neuromorphic spike based processor can be found in \cite{10.3389/fnins.2016.00035}. 

\section{Background}
\label{sec:background}
This section provides an overview of event-based vision sensors and their principle of operation. The ARMS flow algorithm used as the basis of this research is discussed in further detail. Finally, a brief discussion of the SoC design methodology and tools used in this research is provided. 

\subsection{Event-Based Vision Sensors}

Unlike traditional cameras, which sample pixel intensity at a fixed, synchronized frame rate, event-based vision sensors record asynchronous pixel events. These events encode temporal log intensity contrast at a pixel as either an ``on'' or ``off'' event for increasing and decreasing intensity over time respectively \cite{Gallego2020Event-basedSurvey}. The sensor outputs events using address-event representation (AER), where each event packet includes the $x$ and $y$ coordinates of the event pixel, the event time $t$, and the event polarity $p$ \cite{Boahen2000Point-to-pointEvents}.

The operating paradigm of event-based vision sensors provides multiple advantages over traditional cameras such as high dynamic range and high temporal resolution \cite{Gallego2020Event-basedSurvey}. High dynamic range allows for detection and tracking of objects in extreme lighting conditions where traditional cameras would be saturated and unable to detect objects. The high temporal resolution of event-based vision sensors is of particular interest for the optical flow application. The microsecond precision of the sensors allows for the possibility of highly accurate optical flow estimates even when objects are moving rapidly through the scene.

This research uses data recorded with a variety of event-based vision sensors with resolutions ranging from 240$\times$180 $px$ to 640$\times$480 $px$ \cite{Brandli2014ASensor,Posch2011ACDS,Son2017ARepresentation}. Event cameras often provide lower resolution than traditional cameras, with early versions having resolutions lower than the ones used in this research \cite{Lichtsteiner2008ASensor}. However, as the technology has matured, higher resolution sensors such as the 1280$\times$720 $px$ sensor presented in \cite{Finateu20200Pipeline} have been developed. Some event-based vision sensors are also able to record grayscale intensities either as synchronous images or asynchronous events along with the temporal contrast event outputs. The calculation of optical flow using ARMS, however, only requires the use of temporal contrast events, therefore grayscale data is not considered.

\Figure[t]()[width=0.99\columnwidth]{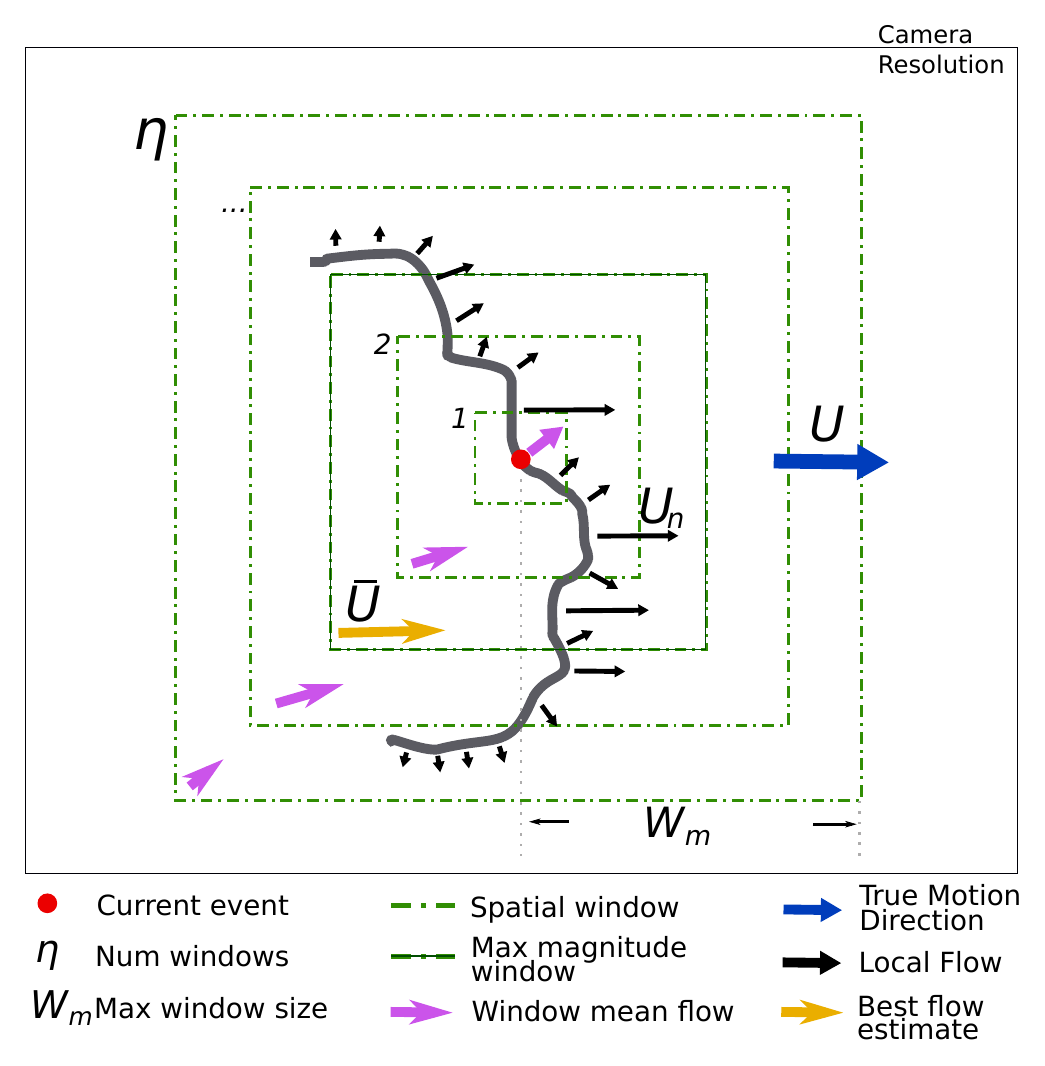}
{\textbf{The principle of aperture robust optical flow computation from a set of local optical flows estimated from increasingly large spatial regions of interest using the ARMS framework. The principle is to determine the spatial window size around the event that has the largest average magnitude of local flow estimates ($U_{n}$). Once found, the average local flow vector ($\bar{U}$) in that maximum magnitude window is used as best estimate for the true optical flow of the event. The selection of the maximum magnitude window corresponds to the selection of the best aperture size for flow computation.} \label{fig:ARMS_Explain}}

\subsection{Aperture Robust Multi-Scale Flow }
\label{sec:multiscale-pooling}
Optical flow computations in general use the movement of pixel-based information such as intensity or events to calculation direction of motion. These true direction estimates, however, are hard to estimate due to the ``aperture problem'' which arises when the flow computation is performed only on a section of the object. To reliably compute the true direction of motion the flow algorithm needs information about the motion of the whole object. This requirement, however, can lead to a computational bottleneck as it requires additional steps such as segmentation or clustering of the different objects in the scene. Recently, we proposed a novel method, called ARMS (Aperture Robust Multi-Scale flow)~\cite{Akolkar2020Real-timeFlow}, which overcomes these problems in an event-by-event, unsupervised manner that eliminates the need for additional computations. While the details may be found in \cite{Akolkar2020Real-timeFlow}, for readability and completeness, we provide, in this section, a brief description of the basic principles behind the working of this method.

Fig.~\ref{fig:ARMS_Explain} shows the operational principle of the ARMS method. Let us consider a contour shown in gray moving in a horizontal direction indicated by the blue arrow with the true-flow velocity of \textbf{${U}$}. For each event generated by this moving contour, we can compute the local flow (\textbf{${U_n}$}) using the derivative of the surface of events. The direction of this local flow is always orthogonal to the local tangent of the contour and typically does not reflect the true direction of motion. However, the true flow and local flow are related based on the orientation of this edge and the true flow direction as shown in \eqref{eq:norm-true-flow}. Interestingly, as the orientation of edge at which local flow is computed becomes orthogonal to the true-flow direction, the local-flow magnitude goes to its maximum value and is equal to the true-flow magnitude, as the value of $cos(\theta)$ goes to one. 

\begin{equation}
    \mathbf{U_n} = |\mathbf{U}|\cos(\theta)
    \label{eq:norm-true-flow}
\end{equation}

In ARMS, we used this principle to show that it is possible to find a spatial neighborhood window around an event, in which maximizing the average local-flow magnitude is equivalent to minimizing the error difference between average local-flow direction and the true-flow direction. This operation can be summarized in \eqref{eq:arms-error-min} where $k$ is the different spatial neighborhood sizes. This means that the minimization problem in \eqref{eq:arms-error-min} can be used to search for the best spatial neighborhood size. This neighborhood size corresponds to the best ``aperture size'' for the flow computation based on the different objects present in the scene, which is determined without requiring any a-priori knowledge of the scene itself, thus, operating in a fully unsupervised manner while performing the flow computation event-by-event.

\begin{equation}
    \underset{k}{\argmin}(E) = \underset{k}{\argmin}(|\mathbf{U}| - \overline{|\mathbf{U}_n|}) \equiv \underset{k}{\argmax}(\overline{|\mathbf{U}_n|})
    \label{eq:arms-error-min}
\end{equation}

The true flow estimate from this spatial window is then given as the average of all the recent local-flow vectors computed within this window.

The visualization in Fig.~\ref{fig:ARMS_Explain} shows the ARMS realization of this windowing strategy, known as multi-scale pooling. The event at which the true flow is being computed is shown as the red dot. The events on the contour around this event also have already computed, local-flow vectors shown as black arrows. The magnitude of these flow vectors (i.e the length of the arrows) varies depending on the contour section orientation w.r.t the true-flow direction. We create $\eta$ windows of increasing size (shown as green dashed rectangles), starting from the smallest window centered around the event, up to the largest window of size $W_n$. For each of these spatial windows we compute the average magnitude ($\overline{\lvert U_n \rvert}$) of all the local-flow vectors within the window (shown as magenta arrows at the bottom left for each window). We then find the window which has the maximum magnitude (shown as green solid rectangle with orange vector in bottom left). The average local-flow vector ($\overline{U}$) in this window is then assigned as the true optical flow for the event.

It may be noted that the theoretical basis for the ARMS flow algorithm holds for any local flow that fulfils the criteria in \eqref{eq:norm-true-flow}. This means that many different existing methods could be used to calculate local flow and the multi-scale search then could be used to correct the flow direction. We reported in \cite{Akolkar2020Real-timeFlow} that the ARMS method performed considerably faster than any existing state of the art method, without need for specialized computational hardware such as a GPU. However, for some datasets and scenarios, real-time performance could not be achieved due to the deluge of data. The performance of the original ARMS algorithm is constrained by two primary factors: (1) The ARMS algorithm requires computation of the average local-flow vectors on all events within each window leading to repetitive averaging as the larger windows already encompass the events of the smaller neighborhoods and (2) the averaging considers all pixel locations in the $\eta$ windows even if no new recent events occurred at these pixels. This means that the computational complexity of ARMS depends on both the maximum search window size ($W_m$) and the number of windows ($\eta$). Further analysis of the ARMS algorithm complexity is presented in Section.~\ref{sec:alg-complexity-analysis}.

To mitigate these issues with ARMS algorithm, in this paper, we propose optimization of the ARMS algorithm to achieve significantly higher throughput. We refer to this new optimized ARMS algorithm as fARMS. The optimization strategy and the new fARMS algorithm is discussed in detail in Sec.\ref{sec:fARMS-opt-alg}. Further, we realized a high-performance parallel implementation in hardware on a Xilinx Zynq-7000 series SoC. This hardware implementation of ARMS is referred to as hARMS and its performance improvements, throughput, accuracy and different parameter considerations are detailed in later sections.

\subsection{System-on-Chip Development}
\label{sec:zynq-soc-platform}

We used Xilinx's Zynq-7000 series SoC to implement the hARMS architecture presented in this paper. Xilinx's Zynq SoC is a hybrid computing platform that couples a traditional ARM processing system (PS) with a programmable logic (PL) FPGA fabric region. This architecture allows for algorithms to be split across both the regions of the SoC to achieve optimal performance. To streamline the development of accelerated applications targeting the Zynq platform, Xilinx provides the Software Defined System-on-Chip (SDSoC) development tool. SDSoC uses high-level synthesis (HLS) to allow hardware designs to be written in C/C++ for rapid development and verification of designs. SDSoC and HLS were used in the development of hARMS to allow for efficient design iteration and increase the configurability of the design.

Specifically, we used the Zynq-7045 SoC, which contains a dual-core ARM Cortex-A9 processor operating at 667 MHz and a Kintex-7 FPGA \cite{XilinxZynqDatasheet}. This device was chosen due to its use in embedded computing applications combined with its large FPGA fabric to allow scaling to large hARMS configurations. hARMS can also be deployed on other Xilinx SoC platforms depending on the desired application, computing environment, and configuration.

\section{Algorithm Optimization} \label{sec:alg-optimizations}
This section outlines the optimizations made to the ARMS algorithm. The redesigned algorithm, referred to as faster ARMS (fARMS), is presented in detail. Complexity analysis is performed for both the ARMS and fARMS algorithms, and results are compared.

\subsection{Optimized Algorithm} \label{sec:fARMS-opt-alg}

\begin{table}[t]
\renewcommand{\arraystretch}{1.25}
\centering
\caption{Algorithm configuration parameters.}
\begin{center}
\begin{tabular}{c|l}
\toprule
\textbf{Parameter} & \multicolumn{1}{c}{\textbf{Description}}      \\ 
\midrule
$W_m$    & maximum window size for multi-scale pooling \\
$\eta$    & number of spatial windows around event \\
$\tau$     & refraction time for flow events in ARMS algorithm \\
$N$ & length of recent flow event buffer \\
$P$  & number of parallel accelerator cores (hARMS only)\\
\bottomrule
\end{tabular}
\end{center}
\label{table:params}
\end{table}

\begin{algorithm}[t]
\caption{fARMS algorithm for true flow.}
\begin{algorithmic}[1] 
    \STATE $RFB[N] \longleftarrow 0$, recent flow-event buffer
    \STATE $next\_idx \longleftarrow 0$, RFB fill index
    \STATE $EDGE[\eta + 1]$, window bin edges
    \STATE 
    \STATE \textbf{1. Initialize Window Edges}
    \FOR{ $win \longleftarrow 0$ to $\eta$}
        \STATE $EDGE[win] = win \cdot (W_{m}/\eta)$
    \ENDFOR
    \STATE
    \STATE \textbf{2. Process Events}
    \FOR{\textbf{each} $event(x,y,t,vx,vy,mag)$}
        \STATE $sums \longleftarrow 0$, holds vx, vy, and mag sum arrays
        \STATE $COUNTS \longleftarrow 0$, window event count array
        \STATE $RFB[next\_idx] = event$ 
        \STATE $next\_idx = (\mathrel{++}next\_idx) \bmod N$
        \FOR{$i \longleftarrow 0$ to $N$}
            \IF{abs($RFB[i].t - event.t$) $ \leq \tau$}
                \STATE \textbf{2a. Window Arbitration}
                \STATE $dx = \text{abs}(event.x - RFB[i].x$) 
                \STATE $dy = \text{abs}(event.y - RFB[i].y$)
                \STATE $d_{max} = \max(dx,dy)$
                \FOR{ $j \longleftarrow 0$ to $\eta - 1$}
                   \IF{$d_{max} \in [{EDGE[j]},{EDGE[j + 1]}[$}
                        \STATE $tag = j$
                   \ENDIF
                \ENDFOR
                \STATE \textbf{2b. Window Averaging}
                \FOR{ $k \longleftarrow 0$ to $\eta - 1$}
                   \IF{$tag \leq k$}
                        \STATE $sums.VX[j] \mathrel{+}= RFB[i].vx$
                        \STATE $sums.VY[j] \mathrel{+}= RFB[i].vy$
                        \STATE $sums.MAG[j] \mathrel{+}= RFB[i].mag$
                        \STATE $COUNTS[j]\mathrel{++} $
                   \ENDIF
                \ENDFOR
            \ENDIF
        \ENDFOR
        \STATE $MAG\_AVGS = sums.MAG/COUNTS$
        \STATE $w_{max} =$ argmax($MAG\_AVGS$)
        \STATE $true\_vx = sums.VX[w_{max}]/COUNTS[w_{max}]$
        \STATE $true\_vy = sums.VY[w_{max}]/COUNTS[w_{max}]$
        \RETURN \textbf{Flow}($true\_vx, true\_vy$)
    \ENDFOR
    
\end{algorithmic}
\label{alg:fARMS}
\end{algorithm}

Table~\ref{table:params} introduces the parameters used to configure the different ARMS algorithms. The configuration of these parameters impacts the performance and accuracy of the design and must be set based upon the performance requirements of a given application. $W_{m}, \eta,$ and $\tau$ are configured for all implementation whereas $N$ is characteristic only of fARMS and hARMS and $P$ is only used for hARMS configuration.

The ARMS algorithm, presented in \cite{Akolkar2020Real-timeFlow}, relies on successive pooling of events in multiple expanding spatial windows. This design results in repetitive computation of averages in regions around the event that are a subset of multiple spatial windows. It also requires searching the whole $(2W_{m})\times(2W_{m})~$ pixels region around the most recent event, regardless of which, or how many, of the pixels in that region have triggered recent events. This inefficiency is introduced by the reliance on a frame of recent events. The use of an event frame does not align with the asynchronous nature of the event stream generated by the sensor, which has no concept of a frame. We therefore abandon the use of an event frame altogether, and present a more efficient design based on the use of a small time history of recent events.

The redesigned algorithm used for fARMS is presented in Algorithm~\ref{alg:fARMS}. The first, and most significant, optimization is the introduction of the Recent Flow event Buffer (RFB). This buffer stores the last $N$ events generated with a valid local flow. Because the ARMS algorithm only requires the recent-flow events from the last $\tau~ms$, no information will be lost as long as the value of $N$ is greater than or equal to the number of valid events in that time window. In fact, the use of the RFB preserves more information than the use of an event frame. This is because the event frame only preserves the most recent event at each pixel, discarding the older event even if it may have fallen into the $\tau~ms$ time window. The RFB however, has no limitation on the number of events per pixel that can be stored because the location of the event is explicitly included for each entry instead of being implicitly encoded in the event frame location. We hypothesize that multiple events at a single pixel within the $\tau~ms$ window are most likely to occur along strong edges in the scene where the local-flow estimates will be most accurate. Because of this, we expect to see, on average, improved true-flow estimates from the fARMS algorithm when compared to ARMS.

While we do expect to observe improved accuracy from fARMS, the primary objective is optimization for improved throughput performance. The use of the RFB also yields significant performance improvement due to the removal of redundant computation and the reduction of the search space. To enable the use of the RFB and achieve this performance improvement, the challenge of determining which windows each event in the RFB falls into needs to be addressed. Unlike the event frame, the RFB maintains no spatial relationship between events, instead just storing the $x$ and $y$ locations of the event. Therefore, we introduce a window arbitration technique to give each event in the RFB a window tag based on its $x$ and $y$ location relative to the current event. First, maximum component distance between the current event and the RFB event is found. Then a tag is assigned based on the pre-computed window bin that the maximum component distance falls into. It is known that an event that falls into a given spatial window will also belong to all larger spatial windows. This means that only $\eta + 1$ unique window tags are needed to encode all possible windows along with the scenario where an event is not included in any windows. Once the windows that an event falls into are determined, the averaging can be performed as shown in part 2b of Algorithm~\ref{alg:fARMS}. With the use of window arbitration and the RFB the algorithm only requires iteration over all of the $N$ events in the RFB as opposed to costly searches over each of the expanding spatial windows.

The true-flow results from fARMS are calculated in the same way as ARMS. The spatial window with the maximum local-flow magnitude is considered the correct window and the averages of the $x$ and $y$ components of local flow in that window are returned as the true-flow result. A comparison of the complexity of both ARMS and fARMS is provided in the following section and accuracy results are discussed in Section~\ref{sec:experiments-and-results}.

\subsection{Complexity Analysis}
\label{sec:alg-complexity-analysis}

The worst-case complexity of both the ARMS and fARMS algorithms is evaluated based on the number of loop iterations required for the true-flow computation for a single event. The number of loop iterations, $n_{ARMS}$, for the original ARMS algorithm in \cite{Akolkar2020Real-timeFlow} is shown in \eqref{eq:ARMS-complexity-summation}.

\begin{equation}
    n_{ARMS} = \sum^{\eta}_{i = 1}{\left(\frac{2W_{m}}{\eta}\right)^2i^2}
    \label{eq:ARMS-complexity-summation}
\end{equation}

Expansion of the summation in \eqref{eq:ARMS-complexity-summation} yields the expression for $n_{ARMS}$ given in \eqref{eq:ARMS-complexity-n}. From \eqref{eq:ARMS-complexity-n} the complexity in terms of loop iterations is derived and shown in \eqref{eq:ARMS-complexity}.

\begin{equation}
    n_{ARMS} = \frac{1}{6}\left(\frac{2W_{m}}{\eta}\right)^2\eta\left(\eta + 1\right)\left(2\eta + 1\right)
    \label{eq:ARMS-complexity-n}
\end{equation}

\begin{equation}
    n_{ARMS} \in O( W_{m}^2\eta )
    \label{eq:ARMS-complexity}
\end{equation}

From \eqref{eq:ARMS-complexity} it can be seen that the complexity of the ARMS algorithm is bounded by $W_{m}^2$ and $\eta$. The squared dependence on $W_{m}$ poses significant challenges for the scaling of the algorithm, especially as the resolution of the sensor increases. To capture the same region of the scene within the spatial windows, $W_{m}$ must scale as the sensor resolution scales, otherwise an insufficient portion of the scene may be considered for evaluating the true flow at an event. This could reduce robustness to the aperture problem and the overall effectiveness of the algorithm. Furthermore, it is later observed in Section~\ref{sec:experiments-and-results} that flow accuracy tends to improve with larger values of $W_{m}$, particularly when there is a single dominant direction of motion in the scene. The necessity of scaling to higher values of $W_{m}$ in many cases means that complexity that scales independent from $W_{m}$, and thus sensor resolution, is highly desirable.

The number of loop iterations required to compute true flow for one event using the fARMS algorithm is shown in \eqref{eq:fARMS-complexity-summation}. From this the overall complexity is derived in \eqref{eq:fARMS-complexity}.

\begin{equation}
    n_{fARMS} = \sum^{N}_{i = 1}{2\eta}
    \label{eq:fARMS-complexity-summation}
\end{equation}

\begin{equation}
    n_{fARMS} = 2N\eta
    \label{eq:fARMS-complexity-n}
\end{equation}

\begin{equation}
    n_{fARMS} \in O( N\eta )
    \label{eq:fARMS-complexity}
\end{equation}

From \eqref{eq:fARMS-complexity} we see that the fARMS complexity is bounded only by $N$ and $\eta$. Therefore, the fARMS algorithm achieves the objective of scaling independent of $W_{m}$. Since both ARMS and fARMS complexity scales with $\eta$, we compare $W_{m}^2$ and $N$ to evaluate the relative scaling of both algorithms. In most cases $W_{m}^2$ is much larger than $N$, meaning fARMS has a much lower run-time complexity. Take for example a benchmark configuration that will be used in Section~\ref{sec:trivial-pattern} where $W_{m} = 320$, $\eta = 4$, and $N = 1000$. When substituting these parameters into \eqref{eq:ARMS-complexity-n} and \eqref{eq:fARMS-complexity-n} we get $n_{ARMS} = 768000$ and $n_{fARMS} = 8000$. In this case the fARMS algorithm results in a 98.96\% reduction in the theoretical complexity of the true-flow calculation when compared to the original ARMS algorithm. While the complexity difference will vary as the values of $W_{m}$ and $N$ are changed, this analysis shows that the fARMS algorithm substantially reduces ARMS computational complexity.

\section{hARMS System Architecture}

The hARMS system architecture describes the hardware realization of the fARMS algorithm for improved performance on embedded platforms. The system architecture was designed to be modular to allow for streamlined realization of various configurations. The result is a flexible acceleration architecture that can be adapted to application-specific needs. Fig.~\ref{fig:high-level-arch} shows this system architecture with each of the main processing modules included.

\Figure[t]()[width=\linewidth]{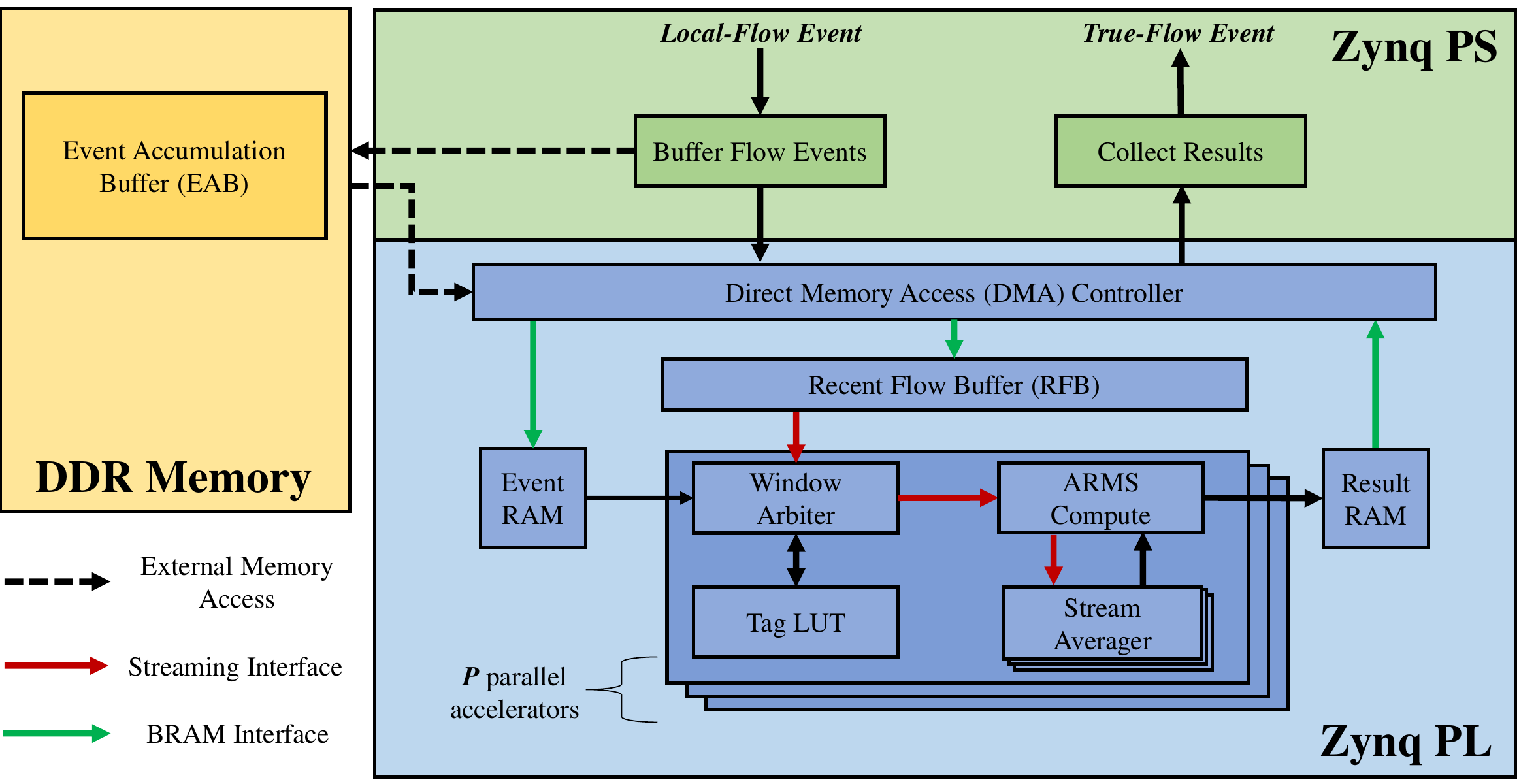}
{\textbf{High level Zynq-SoC acceleration architecture showing the design of hardware ARMS (hARMS). The processing system (PS) and programmable logic (PL) regions, along with the DDR memory shown are hardware components of Zynq platform. DDR memory is used to store the Event Accumulation Buffer (EAB). The PS is used to buffer the incoming local-flow events, initiate memory transfers to the PL, and collect the output true-flow results after computation. Each block in the PL region represents physical hardware implemented in the FPGA fabric. The DMA controls the transfer of data from both DDR memory and the PS to the PL region. The recent flow buffer is a BRAM ring buffer that stores recent local flow results. The window arbiter, tag LUT, stream averager, and ARMS compute blocks are the main computational components of the accelerator, and are duplicated for each of the $P$ parallel accelerators used.} \label{fig:high-level-arch}}

The design is divided across three sections of the Zynq SoC platform---DDR memory, processing system (PS), and programmable logic (PL). The DDR memory is used to buffer flow information and store software application variables. The PS consists of a dual-core ARM Cortex-A9 processor and is used to run the main C++ application that receives local-flow inputs, calls the hardware accelerator, and collects the true-flow results. The PL is used to realize the hARMS accelerator using custom hardware blocks implemented in the FPGA fabric.

The architecture developed for validation and testing computes the local flow in software on the PS. However, the local flow is only considered as an input to the hARMS design and can therefore be computed using any method, including PL acceleration as required by the application. As events are generated and the local flow is computed, the PS is used to accumulate local-flow events and collect the true-flow results. The accumulated events for which a valid local flow exists, and thus true flow can be calculated, are transferred to temporary RAM in the PL fabric using a direct memory access (DMA) controller and a block RAM (BRAM) interface. In addition to being stored in temporary RAM, the accumulated events are also added to the RFB to be used in the processing of future events. The true flow at each accumulated event is then processed in parallel as the RFB is streamed through the $P$ parallel accelerators in the PL region. The processing is performed in a hierarchical order by the window arbitration, stream averaging, and ARMS compute modules. The design of these modules is discussed further in the following sections.

The hARMS design aims to closely match the event-by-event results of the fARMS algorithm. However, for optimization of the hardware implementation, some quantization from full floating point representation is performed. The local-flow results are rounded and represented as 16-bit integer inputs, while the resulting true flow is represented as a 32-bit fixed point value with eight fractional bits. Arbitrary bit width representations are used internally in the design to achieve more resource efficient hardware. Further reduction in the number of bits used to represent data could be considered in some cases depending on the expected range of velocities in the scene or tolerable loss in flow accuracy. We, however, designed the standard hARMS configuration with the goal of being robust to wide variations in scene dynamics, while maintaining accuracy equivalent to the fARMS algorithm.

The architecture includes all the configurable parameters available in the fARMS algorithm, with the addition of $P$ as outlined by Table~\ref{table:params}. Unlike the fARMS algorithm in software, which can define these parameter at runtime, the hardware nature of the hARMS architecture requires configuration before compilation. The value of $P$ is used to specify the number of parallel accelerators to be used in the PL fabric. The high levels of parallelism that can be achieved through the use of this parameter have substantial impact on the design's performance and resource utilization. This impact along with the trade-offs between other parameters will be discussed and analyzed in Section~\ref{sec:experiments-and-results}.

\subsection{Event Accumulation}

Event accumulation is the primary function of the PS in the system architecture. As local-flow events are generated, the flow components and flow magnitude are stored in the Event Accumulation Buffer (EAB). The EAB holds the events for which the true flow will be calculated when the hardware accelerated function is called. The depth of the EAB is equal to the number of parallel accelerators, $P$, included in the design. When the EAB is filled with new events, a DMA transaction is initiated to move the EAB data to the PL region via a BRAM interface. This transaction begins the true-flow calculation process for the events in the EAB. Once the EAB is transferred to the PL, the events are added to the RFB as well as to temporary RAM. The RFB is a BRAM ring buffer of length $N$ that is implemented in the PL fabric and retains its values between calls to the hardware accelerator. The $N$ parameter dictates the number of events stored in the RFB and processed in the PL. The ring buffer allows for new events to be added while replacing the oldest events at the same time. This way the most recent $N$ events are always available for each true-flow calculation. When the true-flow calculations begin, the data stored in the RFB is streamed into the accelerator modules as shown in Fig.~\ref{fig:high-level-arch}. Since the hARMS accelerators are designed for streaming inputs, one value from the RFB can be read each clock cycle, achieving an initiation interval of one and preventing the need for resource-expensive array partitioning. 

The hARMS accelerators are designed to perform asynchronous computation, such that, as the ARMS true flow is being calculated using the FPGA accelerator, the EAB can begin to be filled asynchronously as the previous EAB events are being processed. Buffering events allows multiple true-flow events to be processed simultaneously using the same RFB. This reduces the BRAM required for simultaneous processing of events, while increasing the throughput of the system. Processing multiple true-flow events in this way results in up to $P - 1$ future events being considered for a given event when multi-scale pooling is performed. Because $P$ is typically much smaller than $N$, this artifact of buffering events has no significant impact on the accuracy of the flow estimate as shown by the results in Section~\ref{sec:trivial-dir-est-accuracy}.

\subsection{Window Arbitration}

Multi-scale pooling as introduced in Section~\ref{sec:multiscale-pooling} requires averaging over expanding spatial windows around the pixel location of the incoming event at which true-flow is being calculated. The window or windows that a recent flow event will fall into is dependent on the location of the recent event around which multi-scale pooling is being applied. The hARMS architecture uses the same window arbitration technique introduced by fARMS in Algorithm~\ref{alg:fARMS}. The maximum component distance calculated is used as the input to a hardware lookup table ($tagLUT$), which determines the appropriate window tag. The maximum component distance is equal to $\max(F_{width},F_{height}) - 1$, where the sensor resolution is $F_{width} \times F_{height}~px$. This value could be considered the theoretical maximum value of $W_{m}$, because it is applied outward from the event in all four directions. However, in reality the value of $W_{m}$ is not restricted due to the resolution indifferent design of the hARMS architecture.

As discussed in Section~\ref{sec:fARMS-opt-alg}, there are $\eta+1$ possible window tags that could be assigned to any recent event as it is streamed into the accelerator, with the additional tag representing a recent event that does not belong to any of the defined windows. The tag can then be represented using only $\lceil \log_{2}(\eta + 1) \rceil$ bits in hardware. These tag bits are appended to the input stream, while the $x$ and $y$ coordinate data is removed from the stream, as all required location information is now encoded in the window tag.
 
The window arbiter hardware is fully pipelined to allow one new recent-flow event to be read from the internal input stream generated from the RFB on each cycle. The window edge values are statically declared in the $tagLUT$ module and the window search is fully unrolled. This implementation allows the $tagLUT$ module to achieve an interval and latency of one cycle. This high-performance window arbitration is achieved regardless of the relative order in which events are streamed in. This allows for the simple ring buffer realization of the RFB, significantly reducing the control logic for the RFB and eliminating the need for software-based window arbitration before streaming data to the accelerator.

\subsection{Stream Averaging}

Averaging is the fundamental operation of multi-scale pooling. It involves averaging the values of recent local-flow events within each window. This operation is performed by the stream averaging module implemented in the PL fabric. The stream averager is a modified stream-based FPGA implementation of the averaging performed in the fARMS algorithm. It differs in its streaming nature and modular design, which allows parallelism between multiple instances of the module. As events from the RFB are streamed into the module, they are added to an internal array of window sums, where there is one sum for each of the $\eta$ windows used in the design. This addition, however, is only performed for the sums of the windows that the event falls into. The tag assigned by the window arbitration module is used to select the window sums to which the event value should be added. These steps are outlined further in Algorithm~\ref{alg:averager}. 

\begin{algorithm}[t]
\caption{\textsc{averager} streaming algorithm design.}
\begin{algorithmic}[1] 
    \STATE $ISTREAM$ internal input stream
    \STATE $WIN\_SUMS[\eta]$, array of intermediate sums
    \STATE $WIN\_COUNT[\eta]$, count of events in each window
    \STATE $AVERAGES[\eta]$, array of resulting averages
    \STATE $sEvent$, stream input event holding ($tag, value, valid$)
    \STATE \textbf{1. Compute Window Sums}
    \FORALL{events in $ISTREAM$} \label{alg:averager:line:main-loop}
        \STATE $sEvent \longleftarrow ISTREAM$ \COMMENT{load event from stream}
        \FOR{$idx \longleftarrow 0$ to $\eta - 1$} \label{alg:averager:line:sum-loop}
            \IF{$sEvent_{tag} \leq idx$ and $sEvent.valid$}
                \STATE $WIN\_SUMS[idx] \mathrel{+}= sEvent.value$
                \STATE $WIN\_COUNT[idx] \mathrel{+}= 1$
            \ENDIF
        \ENDFOR
    \ENDFOR
    \STATE \textbf{2. Compute Averages}
    \FOR{$i \longleftarrow 0$ to $\eta - 1$} \label{alg:averager:line:div-loop}
        \STATE $AVERAGES[i] = \dfrac{WIN\_SUMS[i]}{WIN\_COUNT[i]}$
    \ENDFOR
    \RETURN $AVERAGES$
\end{algorithmic}
\label{alg:averager}
\end{algorithm}

As each recent event is added to the appropriate window sums, a count of events that fall into each window is kept. When the entire length of the RFB has been streamed through the averager, the averages for each of the spatial windows are generated by dividing the sum array by the count of events that belonged to the corresponding window. This division occurs once per window for each true-flow event. No checks for division by zero are required because we are guaranteed to have at least one event---the event for which true flow is being calculated---in each window. Because implementation of many dividers is resource intensive, a limit of four hardware dividers per averager module is enforced. These dividers are reused when the number of windows, and therefore divisions, is increased beyond four. Because this operation only occurs once at the end of the processing pipeline for this stage the added latency of pipelined division instead of fully unrolled division only has a limited impact on the overall latency of the stream averaging stage of processing. The constraint of four could be modified to fit specific application needs based on available device resources.

For efficient implementation of Algorithm~\ref{alg:averager} in hardware, the loops on lines \ref{alg:averager:line:sum-loop} and \ref{alg:averager:line:div-loop} are fully unrolled. To facilitate parallel access to the sum and count arrays, they are fully partitioned such that all elements can be accessed and modified concurrently. While the directive is given to fully unroll the division loop on line \ref{alg:averager:line:div-loop}, the unroll factor will, in implementation, be limited by the limit placed on the number of dividers to be instantiated. The compute window sums loop that reads in each of the recent-flow events from the input stream is fully pipelined with an initiation interval of one such that it can read a new event from the input stream once every clock cycle.

\subsection{ARMS Compute Core}

The ARMS compute core is the main functional control block of each hARMS accelerator instantiated. It receives the tagged event stream from the window arbiter. Using the event timestamps included in that stream and the defined value of $\tau$ it performs temporal filtering of the event streams. Any event that occurs more than $\tau$ microseconds before the EAB event under consideration is flagged and not considered when performing multi-scale pooling using the stream averaging blocks.

\begin{algorithm}[t]
\caption{ARMS compute core algorithm for hARMS architecture.}
\begin{algorithmic}[1] 
    \STATE $event$, EAB event of interest ($x, y, t$)
    \STATE $sEvent$, stream input event ($tag, vx, vy, mag, t$)
    \STATE $ISTREAM$, output stream from \textsc{window\_arbiter}
    \STATE \textbf{1. Process Input Stream}
    \FORALL{events in $ISTREAM$} \label{alg:ARMS-Compute:line:main-for}
    \STATE $sEvent \longleftarrow ISTREAM$ \COMMENT{load event from stream}
        \IF{abs($sEvent.t - event.t$) $ \leq \tau$}
            \STATE $valid = \TRUE$
        \ELSE 
            \STATE $valid = \FALSE$
        \ENDIF
        \STATE $VX\_STREAM \longleftarrow (sEvent.(vx,tag), valid)$
        \STATE $VY\_STREAM \longleftarrow (sEvent.(vy,tag), valid)$
        \STATE $MAG\_STREAM\longleftarrow(sEvent.(mag,tag), valid)$
    \ENDFOR
    \STATE \textbf{2. Stream Averaging} 
    \STATE $VX\_AVGS = \textsc{averager}(VX\_STREAM)$ \label{alg:ARMS-Compute:line:Avger1}
    \STATE $VY\_AVGS = \textsc{averager}(VY\_STREAM$
    \STATE $MAG\_AVGS = \textsc{averager}(MAG\_STREAM)$ \label{alg:ARMS-Compute:line:Avger3}
    \STATE \textbf{3. True-Flow Selection}
    \STATE $w_{max} =$ argmax($MAG\_AVGS$)
    \RETURN ($VX\_AVGS[w_{max}], VY\_AVGS[w_{max}]$)
\end{algorithmic}
\label{alg:ARMS-Compute}
\end{algorithm}

The ARMS compute core extracts the three values from the stream that must be averaged: the $x$ component of local flow, $y$ component of local flow, and magnitude of local flow. These values and the window tag are passed to three instances of the stream averaging module in parallel streams. The multi-scale average arrays generated are then collected by the compute core and a maximum search is performed on the flow magnitude average results to find the spatial window with the largest local-flow magnitude. The average $x$ and $y$ components of the local flow in that spatial window are then returned as the true-flow results for the EAB event being processed.

The ARMS compute module functionality described is realized as shown in Algorithm~\ref{alg:ARMS-Compute}. The control flow is modified from that of fARMS to efficiently use streaming interfaces, as well as to capitalize on available parallelism. The process input stream loop is fully pipelined such that one event is read from and written to the window arbiter and averager modules respectively on each clock cycle. The calls to the stream averaging modules in lines~\ref{alg:ARMS-Compute:line:Avger1} to \ref{alg:ARMS-Compute:line:Avger3} are performed in parallel using task-level, dataflow pipelining. Once the stream averaging is completed, the window index, $w_{max}$, corresponding to the maximum local-flow magnitude is found and the true-flow results are returned and stored in temporary RAM before the DMA transfer of the results back to the PS for collection.

\section{Experiments and Results}
\label{sec:experiments-and-results}

We evaluated the presented fARMS and hARMS designs against the same datasets presented in \cite{Akolkar2020Real-timeFlow} for a direct comparison of accuracy and performance between the hardware and software designs. These datasets span a variety of scenes and sensor resolutions, allowing for a detailed investigation of the design space. We also evaluate how resource utilization and performance change when different configurable parameters are modified, and analyze design trade-offs when selecting a hardware configuration. A real-time performance comparison for the different datasets used is also provided to show successful achievement of real-time operation across a variety of visual scenes, event rates, and sensor resolutions. 

The Xilinx ZC706 development board, which includes the Zynq-7045 SoC, was used for all embedded software and hardware benchmarks. The Zynq's FPGA contains 218k LUTs, 437k FFs, 900 DSP slices, and 19.2 Mb of BRAM. Resource utilization will be considered as a percentage of these total resources. Software benchmarks were run on the Zynq's ARM processor using a single core operating at 667 MHz. Compiler optimization was set to -O3 and Xilinx's PetaLinux distribution was used. All hardware configurations used a 200 MHz clock for both the accelerator and the DMA controller.

\subsection{Trivial Pattern}
\label{sec:trivial-pattern}

The same trivial pattern dataset presented in \cite{Akolkar2020Real-timeFlow} was used for evaluation of the developed hARMS accelerator. The dataset was recorded with a qVGA resolution event-based sensor and features a square and bars moving up and down in front of the stationary event camera. This dataset is denoted ``Bar-Square'' data. In the recording, the bars are always moving perpendicular to the true direction of motion, meaning that the ARMS algorithm should achieve an accurate estimate of the true flow in most cases.

This dataset was used to test more than 60 different hardware design configurations. The number of spatial windows ($\eta$), maximum size of the spatial windows ($W_{m}$), and the number of parallel accelerators ($P$) were all varied and the results evaluated. The direction estimation accuracy, throughput, throughput speedup, FPGA resource utilization, and estimated power usage were all collected to evaluate trade-offs in the design configuration selections.

\subsubsection{Direction Estimation Accuracy}
\label{sec:trivial-dir-est-accuracy}
The bar-Square data involves motion of the scene in only one direction---either moving upward or downward. This means that for any of these movements, an ideal optical flow algorithm should output a direction distribution with one peak and with a standard deviation of zero. We use this to quantify the performance of the different implementations of the flow algorithms. Thus, the direction estimation error is quantified as the standard deviation of flow angle results across all the events. Larger standard deviation of angles indicate larger error in correcting the direction of motion from normal to true direction. A low standard deviation indicates that the true flow calculated has only one primary direction of motion, which is what is expected from the dataset. It is important to note that the configuration that provides the best standard deviation for this visual scene, will not necessarily provide the best results in all scenes, but it does allow comparison of accuracy between multiple algorithms and configurations. The values of $W_{m}$ and $\eta$ which provide the optimal results will vary based on sensor resolution and visual scene activity.

Fig.~\ref{fig:qVGA-direction} provides an example of hARMS flow correction when the scene is moving up (top) and down (bottom). The local-flow results are generally normal to the moving edge and are noisy in both magnitude and direction. The corrected hARMS flow results correctly capture the true direction of motion. This behavior is seen in the direction distribution histogram with strong peaks at $90\deg$ and $-90\deg$ as opposed to the local flow, which has multiple erroneous peaks in direction frequency. The hARMS results also show a more uniform magnitude of flow across the sensor frame.

\Figure[t]()[width=.999\columnwidth]{figures/stump2.png}
{\textbf{Bar-Square results are shown for the two directions of motion, up (top) and down (bottom). Local-flow results are shown using red vectors and hARMS output is show in blue. The hARMS results show good direction correction with consistent magnitude. The direction distribution is shown in the polar histogram on the right. The local and hARMS direction distributions have different frequency scales for improved local flow visibility.} \label{fig:qVGA-direction}}

We also evaluated the direction standard deviation results across multiple values of $\eta$ for the ARMS, fARMS, and hARMS algorithms. These results are shown in Fig.~\ref{fig:qVGA-acc-ncores} which uses values of 320 for $W_{m}$ and 1000 for $N$. The value of $N$ was chosen to ensure that all of the true-flow events within the temporal window set by $\tau$ are considered. We observe a significant improvement in direction estimation accuracy for fARMS and hARMS over the original ARMS algorithm. This behavior is a result of the optimizations included in the fARMS algorithm. The use of multiple events at the same pixel within the temporal window, as made possible by the ring buffer realization of the RFB, likely improves performance due to the occurrence of this behavior along strong edges where local-flow estimates are most accurate. The fARMS and hARMS results are almost identical for all window sizes, with only slight variance due to the quantization of inputs to the hARMS accelerator. The value of $P$ has negligible impact on direction estimation accuracy for all tested values of $\eta$. This validates the use of a single input buffer of recent-flow events rather than individual buffers for each of the last $P$ true-flow events.

\Figure[t]()[width=.999\columnwidth]{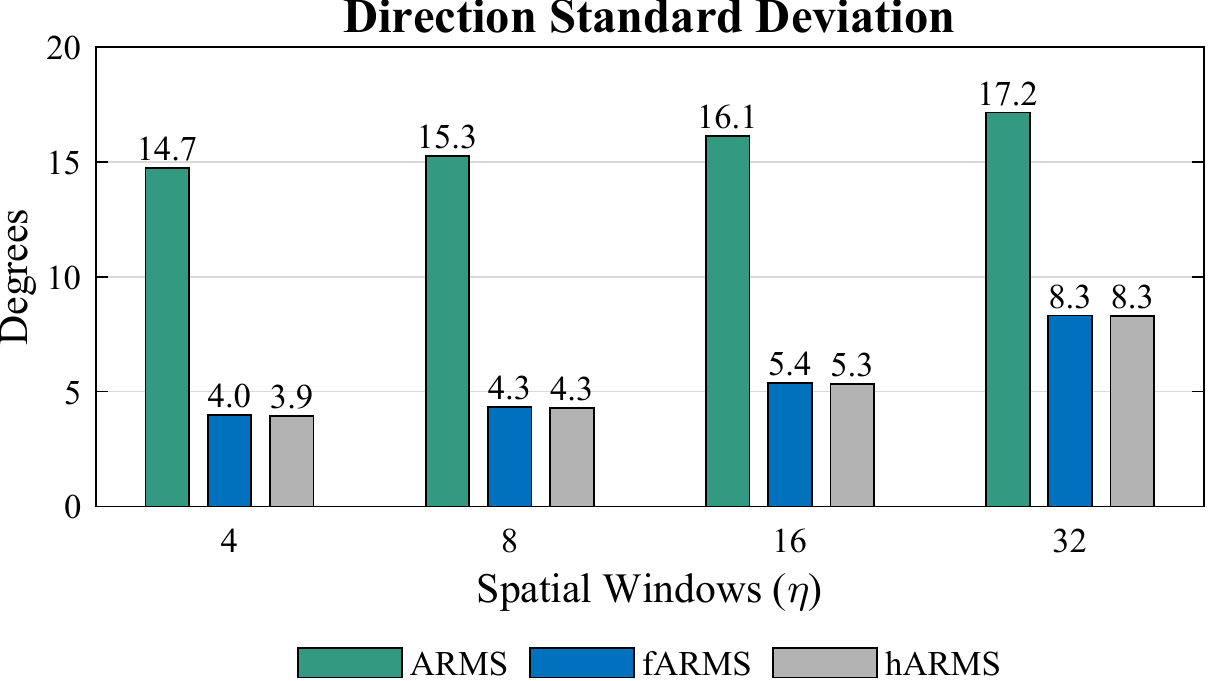}
{\textbf{Flow direction estimates standard deviation for different design configurations. The value of $W_{m}$ is constant at 320 for all results shown and $N$ is fixed at 1000. For this case, fARMS and hARMS results show a significant reduction in standard deviation over the original ARMS algorithm. The hARMS results represent the average of results for all values of $P$.}\label{fig:qVGA-acc-ncores}}

The direction estimation accuracy was also evaluated as a function of RFB length, $N$. While reducing $N$ improves the throughput of the design, it also reduces the number of recent local-flow estimates that can be used for estimating the true direction. Because of this reduction in data, the standard deviation of angle estimates increases as the buffer length is reduced. This behavior is shown in Fig.~\ref{fig:qVGA-acc-bl}. There is also a certain point beyond which increasing $N$ will not improve accuracy because all additional events will be filtered by the temporal window constraint. This is seen in Fig.~\ref{fig:qVGA-acc-bl} between buffer lengths of 1000 and 2000 where there is no change in accuracy. Despite the increase in standard deviation as $N$ decreases, the accuracy was below the of the original software ARMS algorithm. This proves that even with a significant reduction in the amount of data considered, the hARMS and fARMS designs can achieve accurate results.

\Figure[t]()[width=.999\columnwidth]{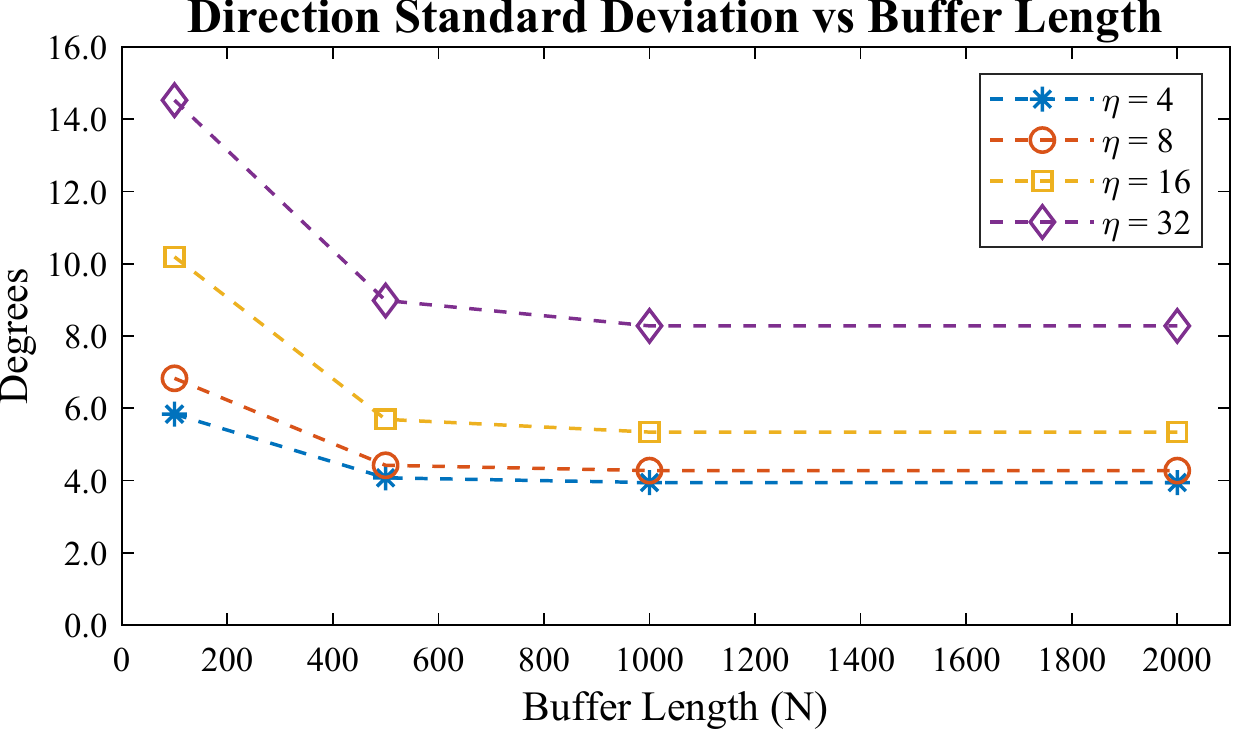}
{\textbf{hARMS standard deviation results for different values of $\eta$ as $N$ changes. Each point represents the average standard deviation across all values of $P$ tested at that configuration. As $N$ increases, the standard deviation decreases until it reaches its minimum.} \label{fig:qVGA-acc-bl}}

\subsubsection{Throughput}
\label{sec:trivial-throughput}
Design throughput is measured in true-flow events per second (evt/s). As many events will not have viable local-flow results such as those generated due to noise, the event rate coming from the sensor could be significantly higher than the maximum design throughput without overwhelming the design. However, we consider only the throughput of the hARMS design after filtering and prepossessing of events occurs. This allows us to evaluate the worst-case scenario where every event generated by the sensor is a valid true-flow event. In Section~\ref{sec:perf-comparison}, a real-time evaluation of the hARMS architecture is provided, which includes information regarding total events and true-flow events in each dataset considered.

Fig.~\ref{fig:qVGA-throughput} shows the throughput results for fARMS and hARMS with varying $P$ for different number of spatial windows ($\eta$). The results shown were generated for a selected benchmark configuration where $W_{m} = 320$ and $N = 1000$. The figure shows that the throughput of the fARMS software design decreases as the number of spatial windows is increased. However, as expected, the throughput for each hardware configuration is nearly constant across varying numbers of spatial windows, with only small decreases as the number of windows increases. This behavior is due to the streaming architecture and fully unrolled window searching implemented in hardware. The small decreases in throughput that are observed as $\eta$ increases are a result of the number of dividers being limited to four per averager. The highest throughput achieved with $N = 1000$ is 1.21 Mevt/s when $\eta = 4$ and $P = 24$.

\Figure[t]()[width=.999\columnwidth]{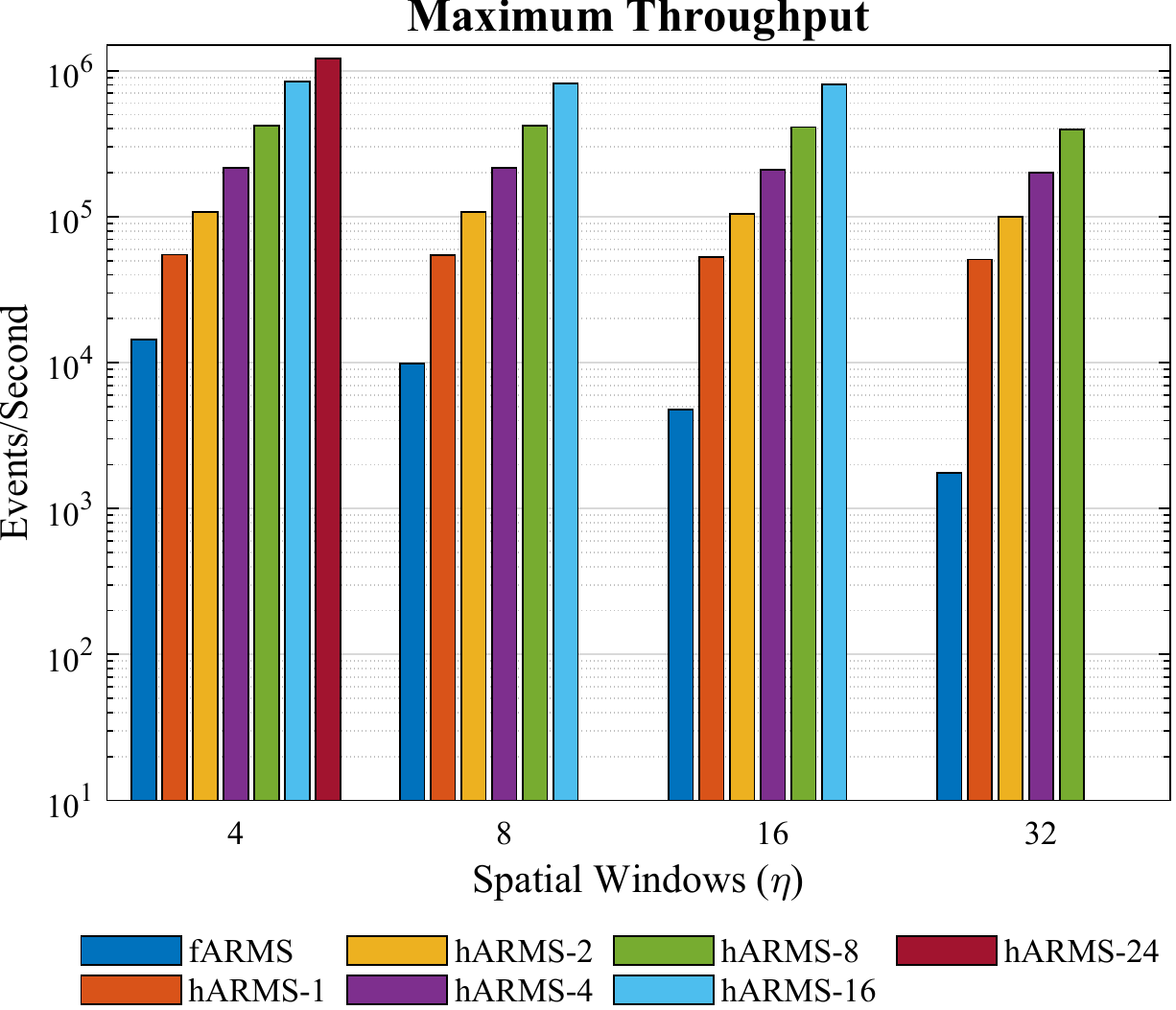}
{\textbf{Maximum throughput results using qVGA Bar-Square dataset. Hardware configurations are denoted as hARMS-$P$. The throughput of the hARMS design increases as $P$ is increased. $W_{m}$ is equal to 320 in all cases and $N$ is 1000 for all cases.}\label{fig:qVGA-throughput}}

Higher throughput speedups were achieved for larger values of $\eta$ due to the decreasing performance of the fARMS baseline with large numbers of spatial windows. That poor performance results from the need for sequential iteration through the windows in software. The parallel nature of hARMS and streaming architecture eliminate this bottleneck and allow for substantial speedup over the software baseline. As shown in Fig.~\ref{fig:qVGA-speedup}, speedup ranges from 4.6$\times$ for $P = 1$ and $\eta = 4$ to 269.2$\times$ for $P = 8$ and $\eta = 32$.

\Figure[t]()[width=.999\columnwidth]{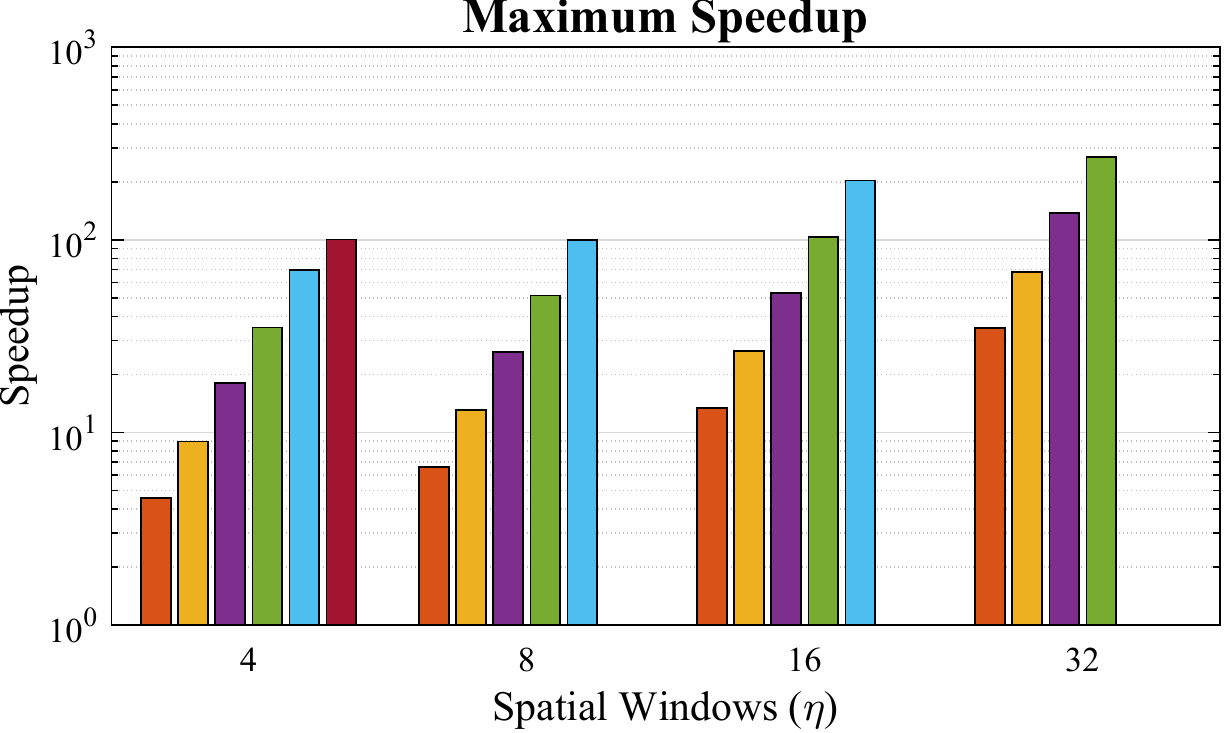}
{\textbf{Maximum speedup results using Bar-Square dataset. The speedup is measured over the optimized fARMS design outlined in Section~\ref{sec:alg-optimizations}. $W_{m}$ is equal to 320 in all cases and $N$ is 1000 for all cases.}\label{fig:qVGA-speedup}}

The value of $W_{m}$ only affects the hardware implementation of the predetermined edges in the $tagLUT$ module, and therefore has no impact on the latency or throughput of the design. The final parameter that has significant impact on the throughput of the design is $N$. Its value was set at 1000 for the primary benchmark because analysis of the Bar-Square dataset showed that a RFB with length 1000 is sufficient to capture all the recent-flow events within the $\tau$ temporal window of 5 $ms$ used. However, in some cases the length of the RFB will need to be larger or smaller based on use case and desired throughput and accuracy. Fig.~\ref{fig:qVGA-throughput-bl} shows the relationship between $N$ and the throughput of one hARMS accelerator core. When $N$ is greater than 1000, the expected trend is observed with throughput decreasing as $N$ increases. Throughput decreases slightly as $\eta$ increases for a given value of $N$. This is caused by the limitation of four hardware dividers per averager in the hARMS core, making repeated calls to division resources necessary for higher numbers of windows. At smaller values of $N$ we observe unexpected behavior, with all configurations having nearly identical flat values for throughput. This is due to the latency of the data transfer between the PS and PL regions, which for small buffer lengths becomes the dominant latency during the hardware function execution. This latency could be removed by providing a direct connection between the event-based vision sensor and the PL region of the SoC. This would result in a significant increase in throughput across all configurations, but it would limit the versatility and configurability of the design as well as require an FPGA implementation of local flow. For these reasons it was not included in the hARMS architecture.

\Figure[t]()[width=.999\columnwidth]{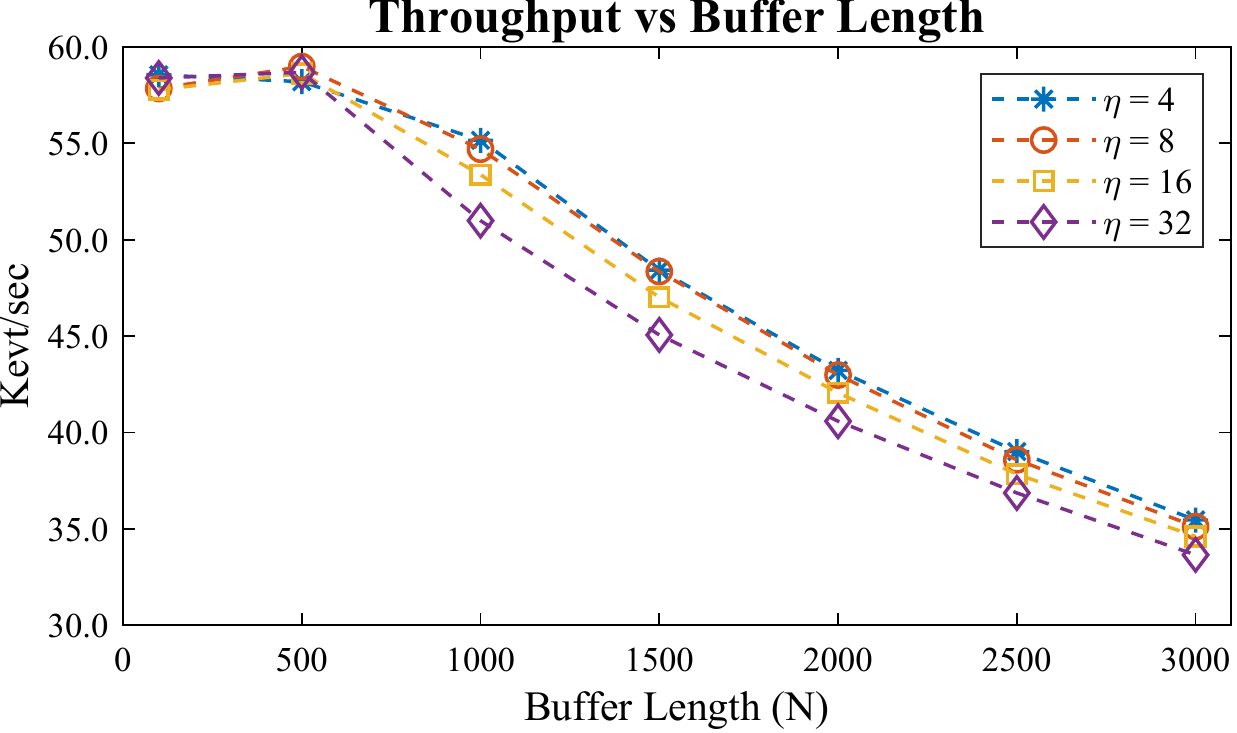}
{\textbf{hARMS throughput measured using Bar-Square dataset for varying buffer lengths and numbers of windows. $P$ is fixed at 1 for all cases.}\label{fig:qVGA-throughput-bl}}

\subsubsection{Resource Utilization}
\label{sec:trivial-fpga-utilization}
FPGA resource utilization is evaluated based on the implemented design for each configuration. LUT and FF usage are of primary interest in evaluating how the design scales as parameters change. BRAM utilization is evaluated in relation to the value of $N$, while DSP utilization remains at zero across all configurations. BRAM usage is independent of all parameters except $N$ as it is only used for the data motion network and the RFB. The volume of data transferred by the DMA controller changes with $P$, this however does not impact the BRAM usage required for the interface meaning that BRAM scales only with $N$.

The LUT and FF usage as a function of $P$ is shown in Fig.~\ref{fig:qVGA-LUT} and Fig.~\ref{fig:qVGA-FF} respectively. Both resources show linear scaling as the value of $P$ increases. The rate of scaling is dependent upon the number of spatial windows used. Table~\ref{table:resources} shows the per parallel accelerator core resource usage for each of the values of $\eta$ tested. This linear scaling is expected due to the limited opportunity for resource sharing between accelerator cores. Only the DMA, event RAM, and result RAM resources can be shared between accelerator cores.

\Figure[t]()[width=.999\columnwidth]{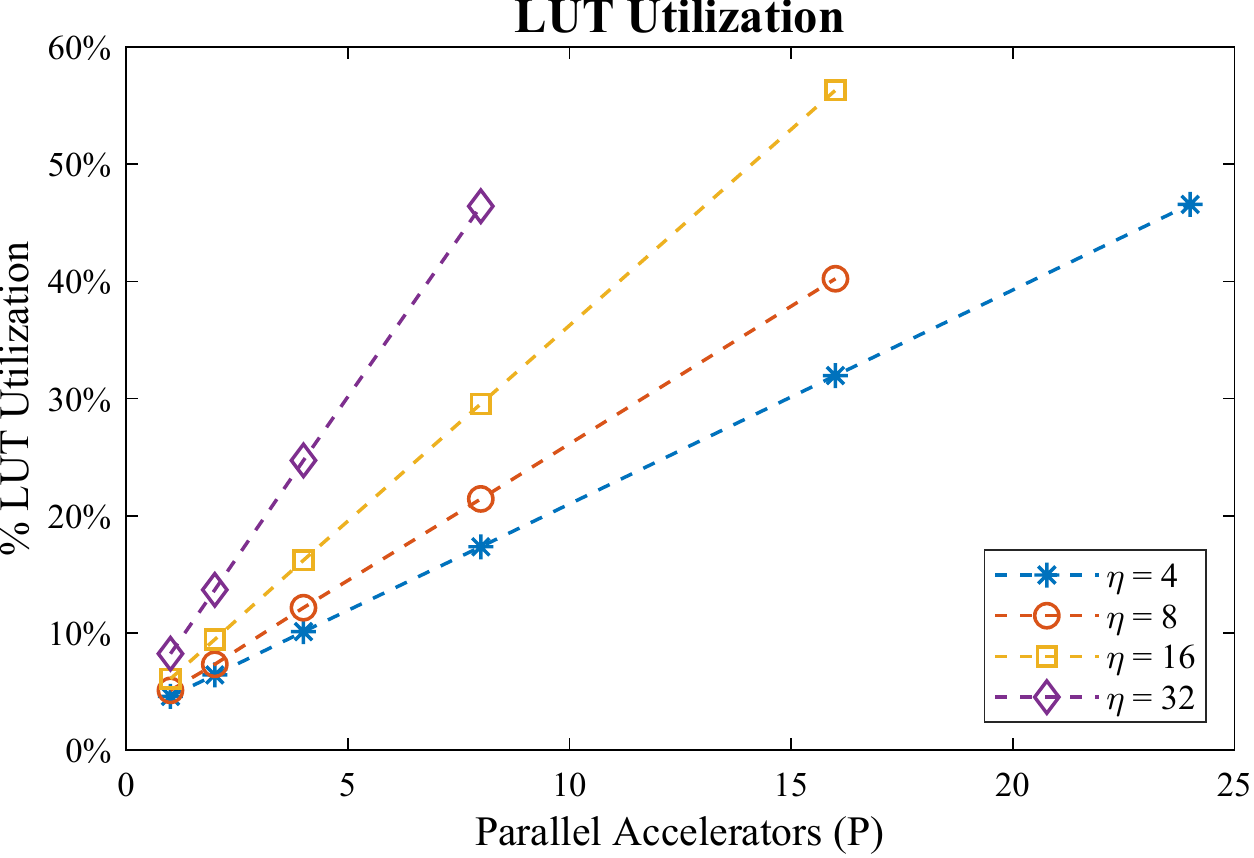}
{\textbf{LUT utilization vs $P$ for different number of spatial windows. Utilization is represented as a percentage of the 218600 total LUT available on the Zynq-7045 SoC used. }\label{fig:qVGA-LUT}}

\Figure[t]()[width=.999\columnwidth]{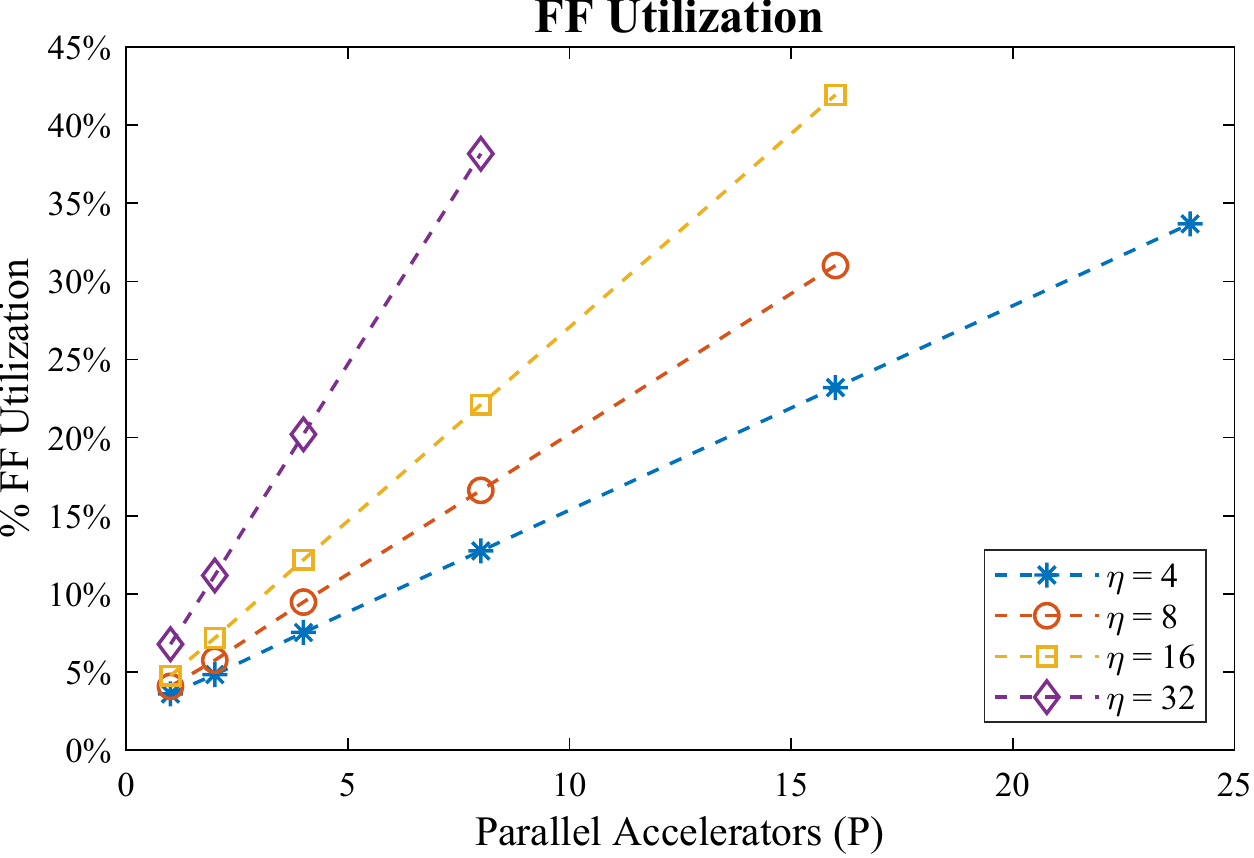}
{\textbf{FF utilization vs $P$ for different number of spatial windows. Utilization is represented as a percentage of the 437200 total FF available on the Zynq-7045 SoC used.}\label{fig:qVGA-FF}}

\begin{table}[b]%[b]
\caption{\textbf{Resource usage per accelerator core for different values of $\eta$ implemented on Zynq-7045 SoC.}}
\centering
\begin{tabular}{@{}ccc@{}} 
\toprule
\textbf{Spatial Windows ($\eta$)} & \textbf{LUT/core} & \textbf{FF/core}\\ 
\midrule
4  & 3959 (1.81\%)  & 5704 (1.30\%)  \\
8  & 5073 (2.32\%)  & 7815 (1.79\%)  \\
16 & 7291 (3.34\%)  & 10848 (2.48\%) \\
32 & 11853 (5.42\%) & 19627 (4.49\%) \\
\bottomrule
\end{tabular}
\label{table:resources}
\end{table}

BRAM utilization as a function of $N$ is shown in Fig.~\ref{fig:qVGA-BRAM}. It shows a baseline utilization of 3.5\%, which is required for the DMA network and a relatively small RFB. As the buffer length increases the BRAM utilization increases. The increase takes on a step-like characteristic because although the absolute size of the RFB increases linearly with $N$, a BRAM block can be considered utilized without its whole memory capacity being used. Overall, BRAM usage is very small compared to the LUT and FF utilization. The hARMS architecture is not dependent on storing an event frame with the same size as the sensor resolution like most existing hardware and software designs. This results in more efficient use of memory in the PL fabric as only the most relevant events are stored.

\Figure[t]()[width=.999\columnwidth]{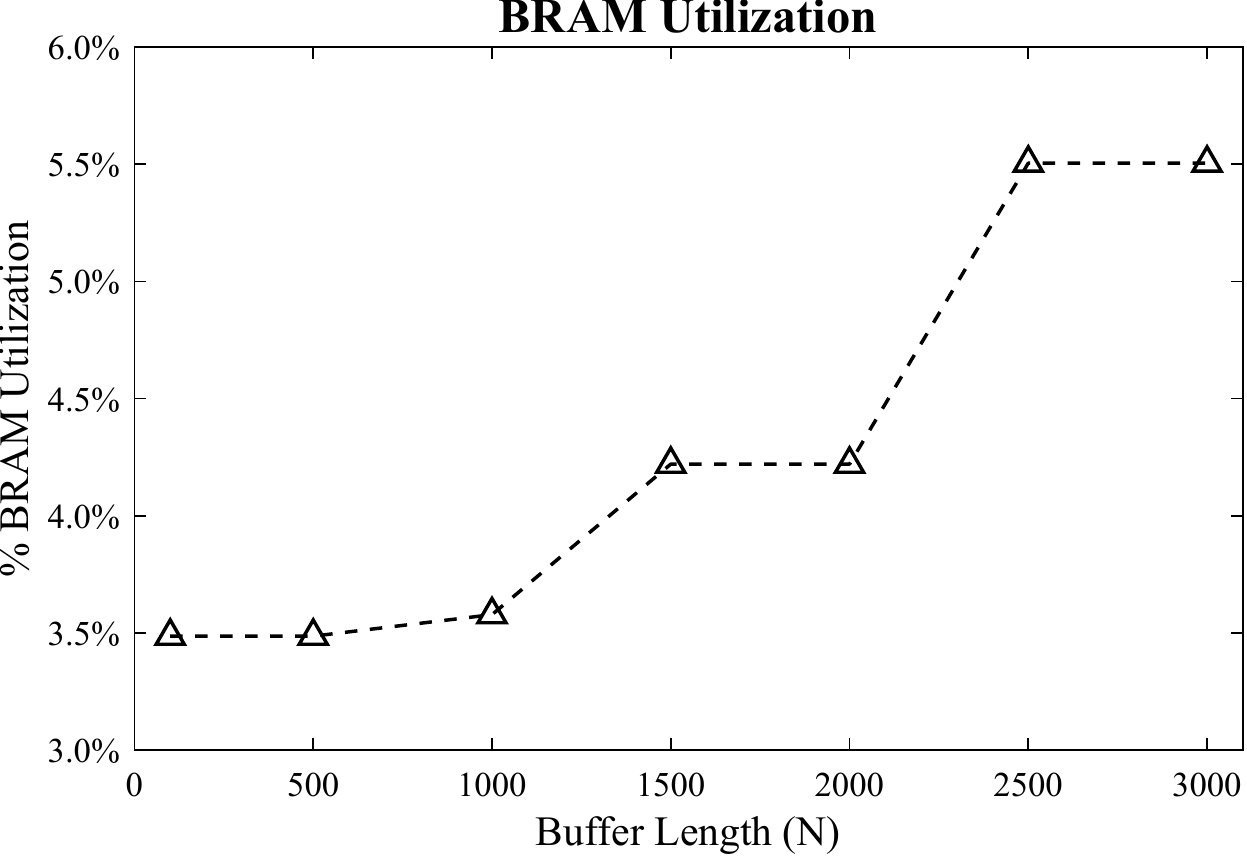}
{\textbf{BRAM utilization vs $N$ for different number of spatial windows. Utilization is represented as a percentage of the 545 total 36Kb BRAM blocks available on the Zynq-7045 SoC used.}\label{fig:qVGA-BRAM}}

\subsubsection{Estimated Power}
\label{sec:trivial-estimated-power}
Power limitations often accompany embedded systems. One advantage of neuromorphic hardware is that it often operates at low-power consumption. To maximize the benefit of sensor power efficiency, it is important to consider the processing power requirements as well. Fig.~\ref{fig:qVGA-est-power} shows how the estimated dynamic power requirements of the FPGA change with the value of $P$. These power estimates are generated from Xilinx's Vivado design tool. Like FF and LUT resource utilization, estimated power consumption scales linearly with $P$.

Overall dynamic power consumption ranges from 0.305 to 1.310 Watts for configurations implemented. Dynamic power for low values of $P$ is dominated by clocking and clock generation which accounts for 200 $mW$ or more of dynamic power depending on the number of windows used. In some cases, this power consumption could become prohibitive for power constrained embedded platforms. To address this concern, steps could be taken to improve power efficiency such as reducing the size of $P$ and $\eta$. Reducing the hARMS operating frequency could also yield reductions in power, however at the cost of reducing the throughput of the design. For this research, we chose to focus on optimization of the design for latency and throughput.

\Figure[t]()[width=.999\columnwidth]{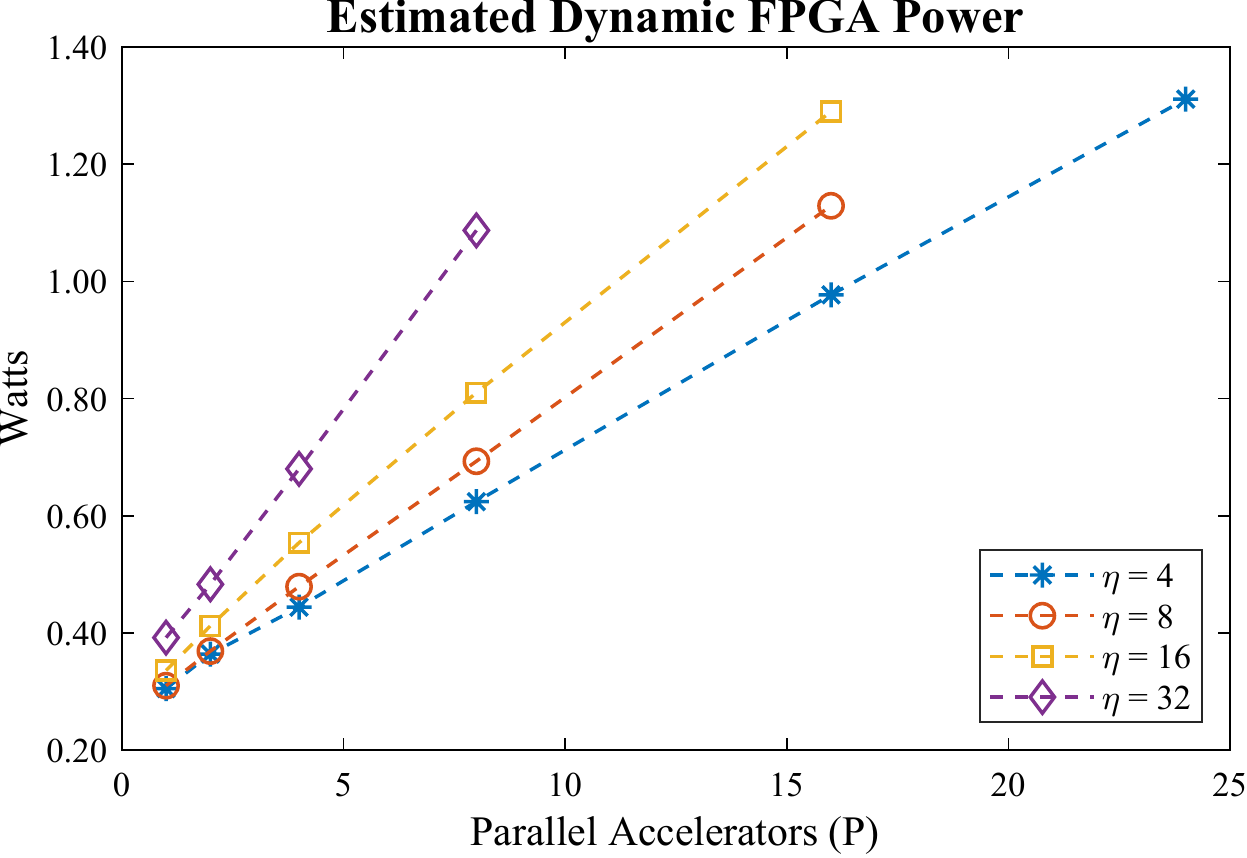}
{\textbf{Estimated dynamic FPGA power consumption vs $P$ for different values of $\eta$. Estimates are obtained from Xilinx Vivado after bitstream generation is completed. Results are for implementation on the Zynq-7045 SoC and do not include the power consumption of the ARM processing system. $N$ is 1000 for all estimates.} \label{fig:qVGA-est-power}}

\section{hARMS on Real-World Datasets}
Based on the results from the Bar-Square dataset, we implemented the hARMS configurations to compute aperture robust optical flow on more complex real-world scenarios such as the DAVIS dataset~\cite{Mueggler2017TheSLAM}, the MVSEC data~\cite{Zhu2018ThePerception} and a VGA resolution recording~\cite{Akolkar2020Real-timeFlow}. The results for these are presented in the next sections. Note that because fARMS and hARMS produce flow results with only small differences due to quantization, only hARMS results will included in the following sections, however, results would be equivalent if using fARMS.

\subsection{Dynamic Rotation Dataset}
\label{sec:davis-dynamic-rotation}

To ensure that the redesigned algorithm used for fARMS and hARMS maintains accuracy in dynamic scenes, the DAVIS dataset presented in \cite{Mueggler2017TheSLAM} was used. This dataset provides the event stream from the 180$\times$240 $px$ resolution DAVIS along with timestamped grayscale images. The dataset also includes inertial measurement unit (IMU) data collected at a rate of 1000 $Hz$, which provides the angular velocity of the camera as the scene is being recorded.

The dataset recording selected is the dynamic rotation scene as it allows for easier mapping from optical flow to the angular velocities from the IMU. In this scene the DAVIS is rotating along its axes while recording an office scene. The dynamic nature of the scene can be seen in Fig.~\ref{fig:DAVIS-qualitative}, which shows the local-flow direction estimates (left) and the hARMS flow direction estimates (right) for three distinct directions of motion. We observe that even in a dynamic scene, the hARMS design performs well, producing the expected true direction of motion as its output, even from noisy local-flow direction estimates. For comparison to the software design study performed in \cite{Akolkar2020Real-timeFlow}, the same algorithm parameters were used, therefore, the hARMS configuration used to process the DAVIS dataset was ($W_{m} = 100$, $\eta = 10$, $\tau = 5ms$, $P = 16$, $N = 1500$). The size of $N$ was intentionally set below the maximum required size to avoid losing events in the temporal window. This was done to show that even with a buffer length of less than half the minimum size to capture all relevant events, the fARMS algorithm and hARMS accelerator can still provide accurate output flow in dynamic scenes. 

The IMU data provided for the dynamic rotation scene was used to quantify the accuracy of the hARMS results in this dynamic scene. The flow velocities in the $x$ and $y$ directions were compared to the angular velocity measurements from the IMU. Both results are shown in Fig.~\ref{fig:dynamic-rotation-results}, which shows high correlation between the hARMS results and the ground truth. The hARMS results achieve a correlation value greater than $R=0.93$ for both the $x$ and $y$ flow estimates.

\Figure[t]()[width=.999\columnwidth]{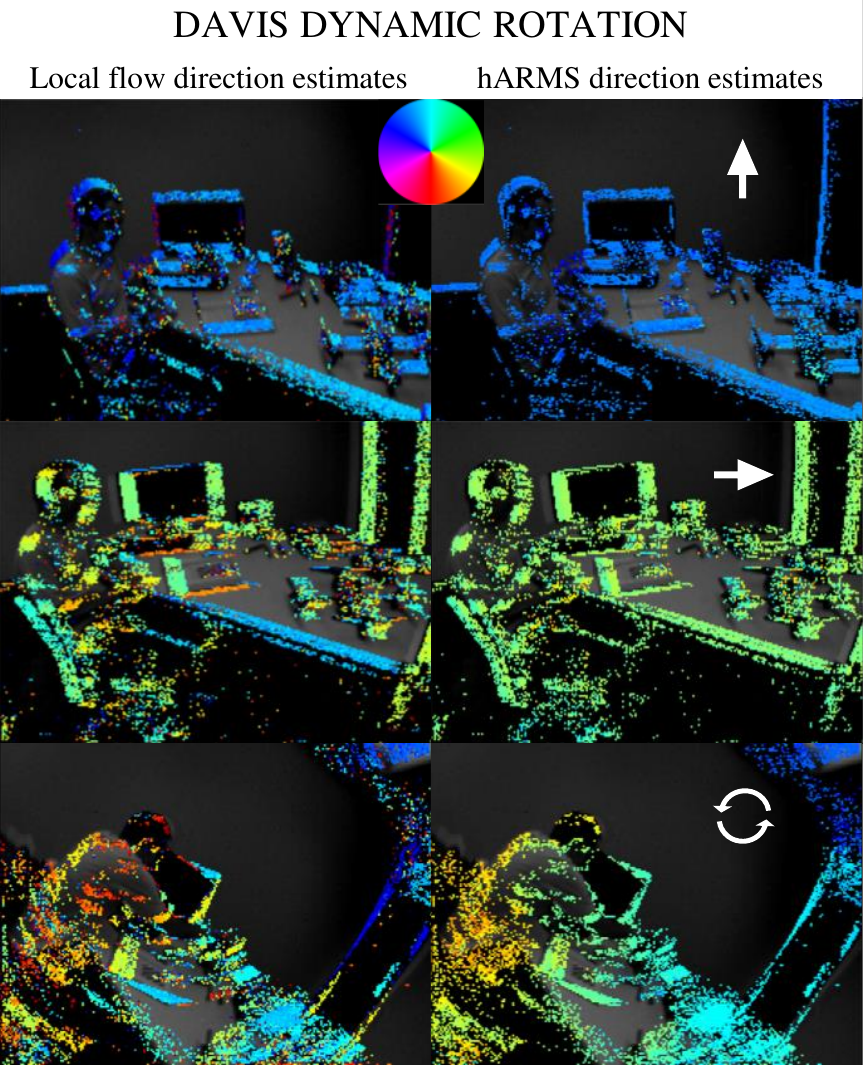}
{\textbf{Qualitative results showing correction of local-flow estimates using the hARMS design. The panels show local and hARMS flow direction estimates for events recorded using DAVIS. The events are overlaid on grayscale images captured using the DAVIS.}\label{fig:DAVIS-qualitative}}

\Figure[t]()[width=.999\columnwidth]{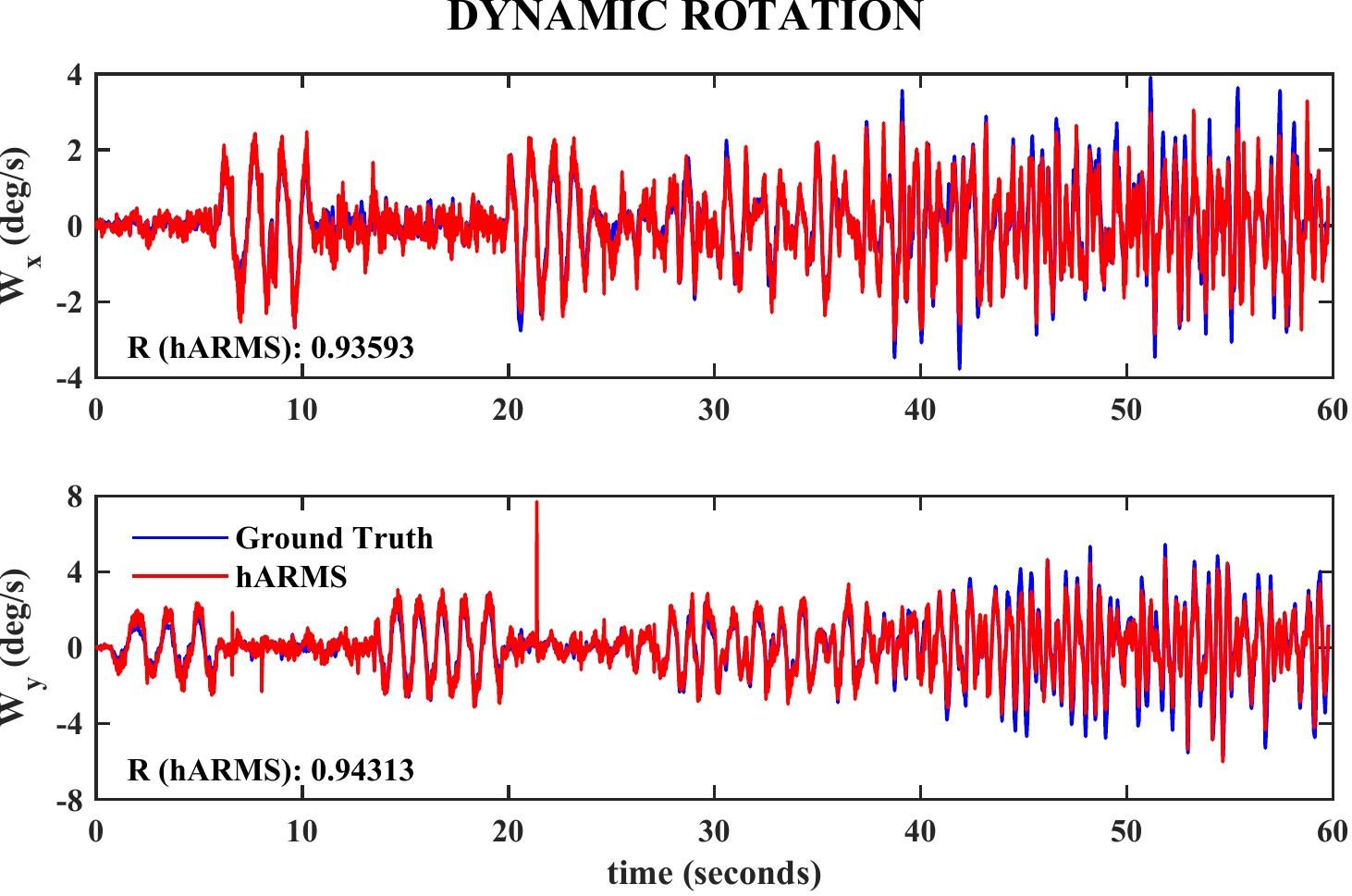}
{\textbf{Comparison of hARMS results with IMU ground truth for dynamic rotation DAVIS dataset. The results show a high correlation between the ground truth angular velocity and the hARMS flow results.} \label{fig:dynamic-rotation-results}}

\subsection{MVSEC Dataset}

For further verification of the hARMS design for optical flow estimation in real-world environments, we applied it to scenes in the Multi Vehicle Stereo Event Camera (MVSEC) dataset \cite{Zhu2018ThePerception}. The MVSEC dataset utilizes an array of sensors to provide dense ground truth optical flow at specified frame intervals. We follow the same approach used in \cite{Akolkar2020Real-timeFlow} to map the asynchronous optical flow outputs from the hARMS implementation to the frame based ground truth. The hARMS configuration used to process the dataset was ($W_{m} = 100$, $\eta = 10$, $\tau = 5 ms$, $P = 16$, $N = 1500$). The hARMS results are plotted with the ground truth for four of the MVSEC dataset recordings as shown in Fig.~\ref{fig:MVSEC-results}. The hARMS results successfully correct the local flow and closely follow the ground truth provided. This is especially true for the indoor flying scenes where the hARMS results closely follow the ground truth throughout the recording. The hARMS flow does struggle more with the outdoor day recording where the hARMS output is much noisier than the ground truth. This is especially noticeable during 22 to 32 second interval when the hARMS results overshoot the $Vy$ ground truth and undershoot the $Vx$ ground truth with high noise in both cases. As was observed in the original software realization presented by \cite{Akolkar2020Real-timeFlow}, this loss of accuracy can occur in areas with poor local-flow estimates or rapid changes in direction for which the ARMS flow cannot quickly correct.

\Figure[t]()[width=.999\columnwidth]{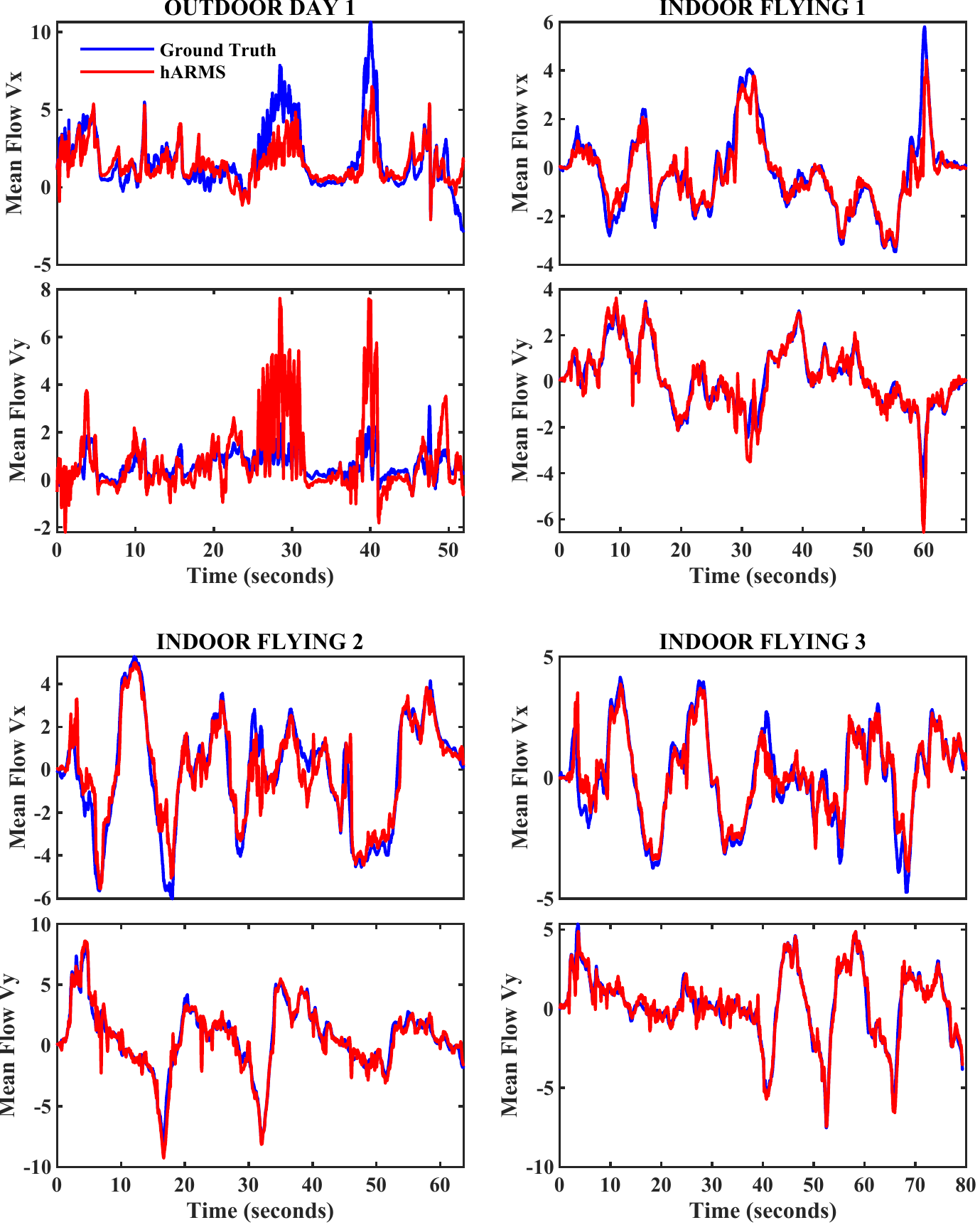}
{\textbf{Comparison of hARMS flow results with the provided ground truth for four MVSEC scenes. hARMS results closely match the ground truth, particularly for the indoor scenes. The hARMS results for the outdoor scene show more variation, but follow the shape of the ground truth well.}\label{fig:MVSEC-results}}

\subsection{Pendulum VGA Dataset}
To verify the performance of the hARMS design for multiple directions of motion and occluded objects within a scene, we use the pendulum dataset introduced in \cite{Akolkar2020Real-timeFlow}. This scene uses a VGA (640$\times$480 $px$) resolution sensor to record two pendulums of the same length oscillating in front of the sensor. One pendulum is placed further from the sensor such that it will appear smaller as it passes behind the other pendulum. For this dataset, we use the following hARMS configuration for processing: ($W_{m} = 50$, $\eta = 5$, $\tau = 5ms$, $P = 16$, $N = 2000$). A smaller value of $W_{m}$ was chosen based on the observation from \cite{Akolkar2020Real-timeFlow} that such selection improves flow results in the presence of occlusion.

Fig.~\ref{fig:pen-qualitative} shows the results of hARMS processing on the pendulum dataset. At 0 ms in the sequence, the pendulums are swinging towards each other and hARMS produces accurate flow estimates for both objects. In the next 40 ms the further pendulum nears the closer and the flow estimates on the leading edge begin to erroneously assume the direction of motion of the closer pendulum. This continues until the further pendulum is completely occluded by the closer pendulum at 80 ms. The same behavior is then observed as the two pendulums separate. The erroneous flow direction estimates are quickly corrected as the distance between the two pendulums increases. This behavior matches that of the original software algorithm and shows that momentary errors in direction due to occlusion are quickly rectified as objects separate in the scene. 

\Figure[t]()[width=.999\columnwidth]{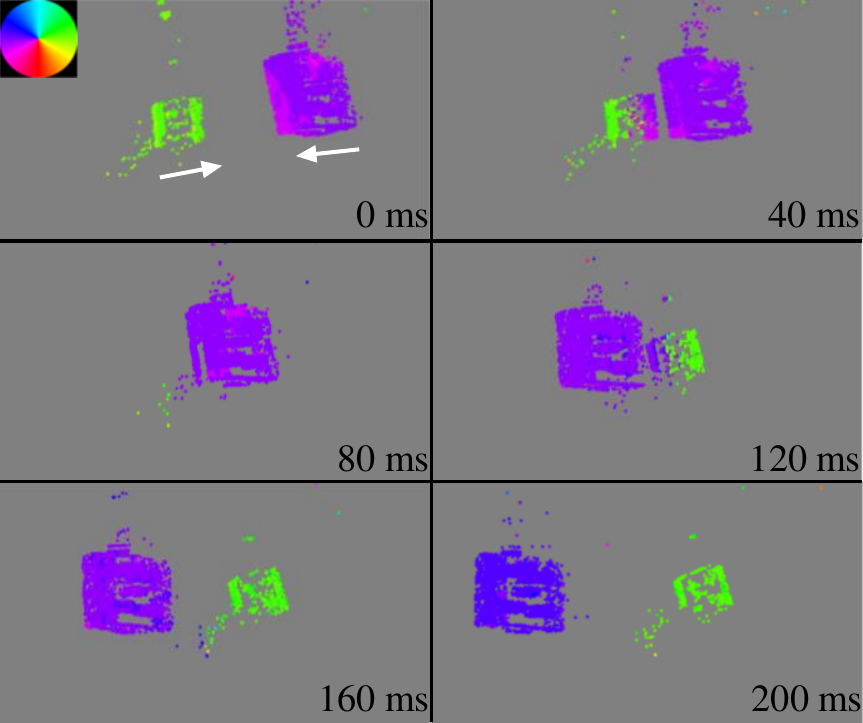}
{\textbf{hARMS results for crossing pendulums at different visual depths. The flow on the further pendulum is distorted as it passes behind the closer pendulum, however, results quickly correct as the objects separate. The event frames shown are accumulated over 20 ms of motion and only the relevant portion of the VGA sensor frame is displayed.}\label{fig:pen-qualitative}}

\subsection{Performance Comparisons}
\label{sec:perf-comparison}

In the following sections we provide real-time performance analysis and comparisons for the fARMS and hARMS designs. Performance evaluations are made for fARMS on both embedded and desktop grade platforms, while hARMS is restricted to its intended use case on embedded platforms. Performance is evaluated across a range of datasets and sensor resolutions.

\subsubsection{Embedded Performance}

\begin{table*}[t]
\centering
\caption{\textbf{fARMS and hARMS throughput performance comparison for various dataset scenes on Zynq-7045 embedded platform. Real-time operation indicated in bold.}}
\begin{tabular}{@{}lcccccccc@{}} 
\toprule
\multicolumn{1}{c}{\textbf{Dataset}} & \begin{tabular}[c]{@{}c@{}}\textbf{Sensor}\\\textbf{Resolution} \end{tabular} & \begin{tabular}[c]{@{}c@{}}\textbf{Total}\\\textbf{Events}\end{tabular} & \begin{tabular}[c]{@{}c@{}}\textbf{True-flow}\\\textbf{Events}\end{tabular} & \begin{tabular}[c]{@{}c@{}}\textbf{Recording}\\\textbf{Duration (sec)}\end{tabular} & \begin{tabular}[c]{@{}c@{}}\textbf{True-flow}\\\textbf{Rate (Kevt/s)}\end{tabular} & \begin{tabular}[c]{@{}c@{}}\textbf{Buffer}\\\textbf{Length}\end{tabular} & 
\begin{tabular}[c]{@{}c@{}}\textbf{fARMS Compute}\\\textbf{Rate (Kevt/s)}\end{tabular} &
\begin{tabular}[c]{@{}c@{}}\textbf{hARMS Compute}\\\textbf{Rate (Kevt/s)}\end{tabular}  \\ 
\midrule
Dynamic Rotation & 240$\times$180 & 71.32e6 & 12.97e6 & 59.77 & 217.00 & 3286 & 5.1  & \textbf{509.1} \\ 
Bar-Square       & 304$\times$240 & 1.25e6  & 530933  & 5.80  & 91.54  & 906  & 15.3 & \textbf{840.0} \\ 
Outdoor Day 1    & 346$\times$260 & 20.0e6  & 3.71e6  & 52.86 & 70.19  & 1267 & 15.3 & \textbf{753.2} \\ 
Outdoor Night 1  & 346$\times$260 & 20.0e6  & 3.80e6  & 53.13 & 71.52  & 2209 & 9.6  & \textbf{612.2} \\ 
Indoor Flying 1  & 346$\times$260 & 12.0e6  & 1.65e6  & 69.02 & 23.91  & 519  & \textbf{43.6} & \textbf{825.9} \\ 
Indoor Flying 2  & 346$\times$260 & 20.0e6  & 2.30e6  & 70.81 & 32.48  & 904  & 26.2 & \textbf{803.9} \\ 
Indoor Flying 3  & 346$\times$260 & 20.0e6  & 2.18e6  & 83.91 & 25.98  & 710  & \textbf{35.4} & \textbf{811.7} \\ 
Pendulum         & 640$\times$480 & 4.82e6  & 618312  & 5.15  & 120.06 & 1853 & 8.9  & \textbf{648.1} \\ 
\bottomrule
\end{tabular}
\label{table:perf-comparison-embed}
\end{table*}

Evaluating the performance of an asynchronous, event-based algorithm for real-time operation can be difficult due to the inherent dependency on the event rate at any moment in time. Scenes with high activity, and therefore high event rates will require higher algorithm processing throughput to achieve real-time operation. For this research we will consider real-time operation to be the case where the fARMS or hARMS compute rate exceeds the true-flow rate of a specific dataset, where true-flow rate is the number of events per second that require a true-flow calculation to be performed.

The dependence of hARMS and fARMS throughput on the buffer length, $N$, adds further difficulty to making performance comparisons across different datasets. Higher throughput can be achieved by reducing the input buffer length, however, this artificially reduces the amount of data available to correct the local-flow direction. To avoid this we select the minimum buffer length required to capture all of the relevant events within the temporal window $\tau$. This requires pre-evaluation of the dataset to determine the buffer length, which is equal to the maximum number of true-flow events within a $\tau = 5 ms$ window. In a real-time application where the events are not pre-recorded, the effective buffer length can be dynamically changed in real time based on the event rate as long as it does not exceed the initial value of $N$ used for hARMS configuration. However, for the purpose of evaluation we utilize the advance knowledge of the dataset to select the worst case buffer length for each dataset and use that value for processing of the entire set.

For evaluation we chose the best performing configuration on the Bar-Square trivial motion dataset which used $\eta = 4$. While the use of four spatial windows may not be optimal in all cases, it proved best for a singular direction of motion and is therefore used in all benchmarks to maintain consistency across datasets. The value of $W_{m}$ does not impact the throughput of the design and can be adjusted based on the sensor resolution and operating environment. For the embedded benchmark testing it was kept at 160. The value of $P$ was set to 16 for all hARMS tests due to its high throughput and reduced resource footprint compared to the 24 core configuration that has high throughput, but can only be utilized when $\eta$ is four. The embedded tests use the same plane-fitting local-flow method as in \cite{Akolkar2020Real-timeFlow} and in the prior experiments for consistency. However, in applications where the local-flow computation could become a bottelneck, we propose the use  of a more computationally efficient method for regularizing the timing of events called Savitzky-Golay plane-fitting presented in \cite{Rueckauer2016EvaluationSensor}.

The results of embedded performance benchmarking are shown in Table~\ref{table:perf-comparison-embed}. Real-time performance is achieved using fARMS for two of the datasets, however, as expected most fall short of real time operation. The hARMS architecture is, however, able to easily achieve real time performance across all datasets tested. The dataset for which real-time performance is most difficult is the dynamic rotation set. This set has by far the highest average true-flow event rate of all datasets evaluated and requires the largest RFB. Despite this, the hARMS compute rate exceeded the true-flow rate by 2.35$\times$ using the defined benchmark parameters. It is also important to note that flow accuracy can be maintained even with a more than 50\% reduction in the buffer length, as shown in Section~\ref{sec:davis-dynamic-rotation}. Leveraging that finding could allow for even more comfortable achievement of real-time processing. The results on the other datasets show that not only can real-time performance be achieved, it can also be achieved with reduced resource utilization depending on the scene. Observing the true-flow rates in Table~\ref{table:perf-comparison-embed} and the throughput results discussed in Section~\ref{sec:trivial-throughput}, it can be seen that real-time operation can be achieved with as little as one or two hARMS cores for most datasets, depending on the configuration. 

Another takeaway from the results shown in Table~\ref{table:perf-comparison-embed} is the independence of fARMS and hARMS throughput from sensor resolution. In fact, the lowest resolution sensor has the slowest compute rate, while higher resolution datasets could be processed much faster. This indicates that the most important factor in determining the fARMS and hARMS throughput performance is the true-flow event rate, which directly impacts the required buffer length. That true-flow event rate is directly and most significantly impacted by the visual scene dynamics. The sensor resolution is a secondary contributor to an increased true-flow event rate and therefore to changes in fARMS and hARMS throughput. 

\subsubsection{Desktop Performance}

The optimizations used in the fARMS algorithm provide a significant decrease in the computational complexity of the algorithm and allowed for real-time performance on embedded platform in some limited cases. Although event-by-event flow results will vary between the original ARMS and fARMS algorithms due to these optimizations, the previous sections have shown that the overall flow results generated by fARMS have equivalent or better accuracy across a wide range of visual scenes when compared to ARMS. Having demonstrated comparable flow accuracy, we now compare the throughput of both software algorithms on a desktop grade platform. As in \cite{Akolkar2020Real-timeFlow}, an Intel E5-1603 processor is used to collect the performance results, which are shown in Table~\ref{table:perf-comparison-desktop}. Both fARMS and ARMS use $W_{m} = 320$ and $\eta = 4$, while the fARMS configuration uses the specified buffer length as determined for the embedded platform tests. Real-time operation is defined in the same way as for the embedded tests.

The results in Table~\ref{table:perf-comparison-desktop} show that real-time performance is achievable with the fARMS algorithm for all but two datasets. The ARMS algorithm, however, is far from real-time in all cases with the specified parameters used. This demonstrates that the fARMS algorithm yields significant throughput performance improvements over the original realization presented in \cite{Akolkar2020Real-timeFlow}. Although these software results show real-time performance in many cases, they still fall significantly short of the throughput obtainable by the hARMS architecture. This shows that the hARMS architecture can yield performance improvements even in non-embedded computing environments. 

\begin{table}[t]
\centering
\caption{\textbf{Original ARMS vs faster ARMS (fARMS) throughput performance comparison for various dataset scenes. Real-time operation shown in bold.}}
\begin{tabular}{@{}lccc@{}} 
\toprule
\multicolumn{1}{c}{\textbf{Dataset}} & \begin{tabular}[c]{@{}c@{}}\textbf{Buffer}\\\textbf{Length}\end{tabular} & 
\begin{tabular}[c]{@{}c@{}}\textbf{ARMS Compute}\\\textbf{Rate (Kevt/s)}\end{tabular} &
\begin{tabular}[c]{@{}c@{}}\textbf{fARMS Compute}\\\textbf{Rate (Kevt/s)}\end{tabular}  \\ 
\midrule
Dynamic Rotation & 3286 & 1.77  & 60.85 \\ 
Bar-Square       & 906  & 2.33 & \textbf{133.439} \\ 
Outdoor Day 1    & 1267 & 1.693 & \textbf{96.527} \\ 
Outdoor Night 1  & 2209 & 2.137 & \textbf{210.036} \\ 
Indoor Flying 1  & 519  & 1.646 & \textbf{207.27} \\ 
Indoor Flying 2  & 904  & 1.775 & \textbf{94.343} \\ 
Indoor Flying 3  & 710  & 1.833 & \textbf{165.234} \\ 
Pendulum         & 1853 & 1.683 & 81.205 \\ 
\bottomrule
\end{tabular}
\label{table:perf-comparison-desktop}
\end{table}

\section{Conclusion}
\label{sec:conclusion}

The asynchronous, high temporal resolution of event-based vision sensors make them ideal for computation of optical flow in the visual scene. This optical flow can then be used for other vision related tasks such as object tracking and collision avoidance. However, for these tasks, the optical flow must be computed in real-time and often on small embedded computing platforms with limited processing power. To capitalize on the benefits of accurate optical flow from event-based vision sensors, the development of fast embedded processing solutions is essential. Although both software and hardware solutions exist to calculate event-based optical flow, no previous solutions have achieved real-time, aperture-robust true-flow calculation on an embedded platform while also utilizing and maintaining the high temporal resolution of the sensor.

In this research we have introduced an optimized event-based optical flow algorithm called fARMS along with a novel hybrid acceleration architecture that fulfills those requirements. The fARMS and hARMS architectures were developed based on the ARMS algorithm presented in \cite{Akolkar2020Real-timeFlow}. The algorithm was modified to be amiable to a more asynchronous, neuromorphic implementation in hardware by discarding the use of an event frame and instead operating asynchronously on only a small history of relevant events. This not only improved flow accuracy, with a decrease in angle standard deviation of up to 73\%, it also allowed for the achievement of real-time processing rates that are independent of sensor resolution. hARMS processing throughput of up to 1.21 Mevt/s was achieved, making it the fastest realization of event-based true-flow demonstrated in literature. Real-time speed was achieved for every dataset considered, with most datasets able to be processed in real time with low resource configurations of the hARMS architecture. Thorough analysis of the design space was performed to determine the relationship between the architecture parameters and the throughput, accuracy, resource utilization, and power of the resulting accelerator. Resource utilization scales linearly with the number of accelerator cores used, with FF and LUT utilization being the main constraints on design scaling. 

Unlike previous works that achieve aperture robust flow (\cite{Liu2017Block-matchingImplementation,Lichtsteiner2008ASensor,Zhu2018EV-FlowNet:Cameras}), the hARMS architecture can operate in real time, fully utilizes the temporal resolution of the sensor, and operates only on temporal contrast events. When compared to the estimated resource utilization in \cite{Liu2019AdaptiveSensors}, the hARMS architecture offers configurations capable of achieving real-time performance with fewer resources. The achieved throughput of 1.21 Mevt/s enables hARMS to nearly match the throughput achieved by the local-flow FPGA implementation in \cite{Aung2018Event-basedFPGA}. Matching this performance means that the design in \cite{Aung2018Event-basedFPGA} could be used to generate the local-flow input to the hARMS architecture. For higher resolution sensors, future work would be needed to overcome the memory limitations present in \cite{Aung2018Event-basedFPGA} if the same high throughput performance is to be achieved by the local-flow implementation. Regardless, it has been demonstrated that the hARMS architecture provides a configurable solution for the real-time computation of aperture-robust event-based optical flow, enabling the use of optical flow for higher level event-based vision on embedded platforms.

\section*{Acknowledgment}
This research was supported by SHREC industry and agency members and by the IUCRC Program of the National Science Foundation under Grant No. CNS-1738783.

\newpage
\vfill

\typeout{} % bug fix
\bibliographystyle{./bibliography/IEEEtran}
\bibliography{./bibliography/references}

\newpage

\begin{IEEEbiography}[{\includegraphics[width=1in,height=1.25in,clip,keepaspectratio]{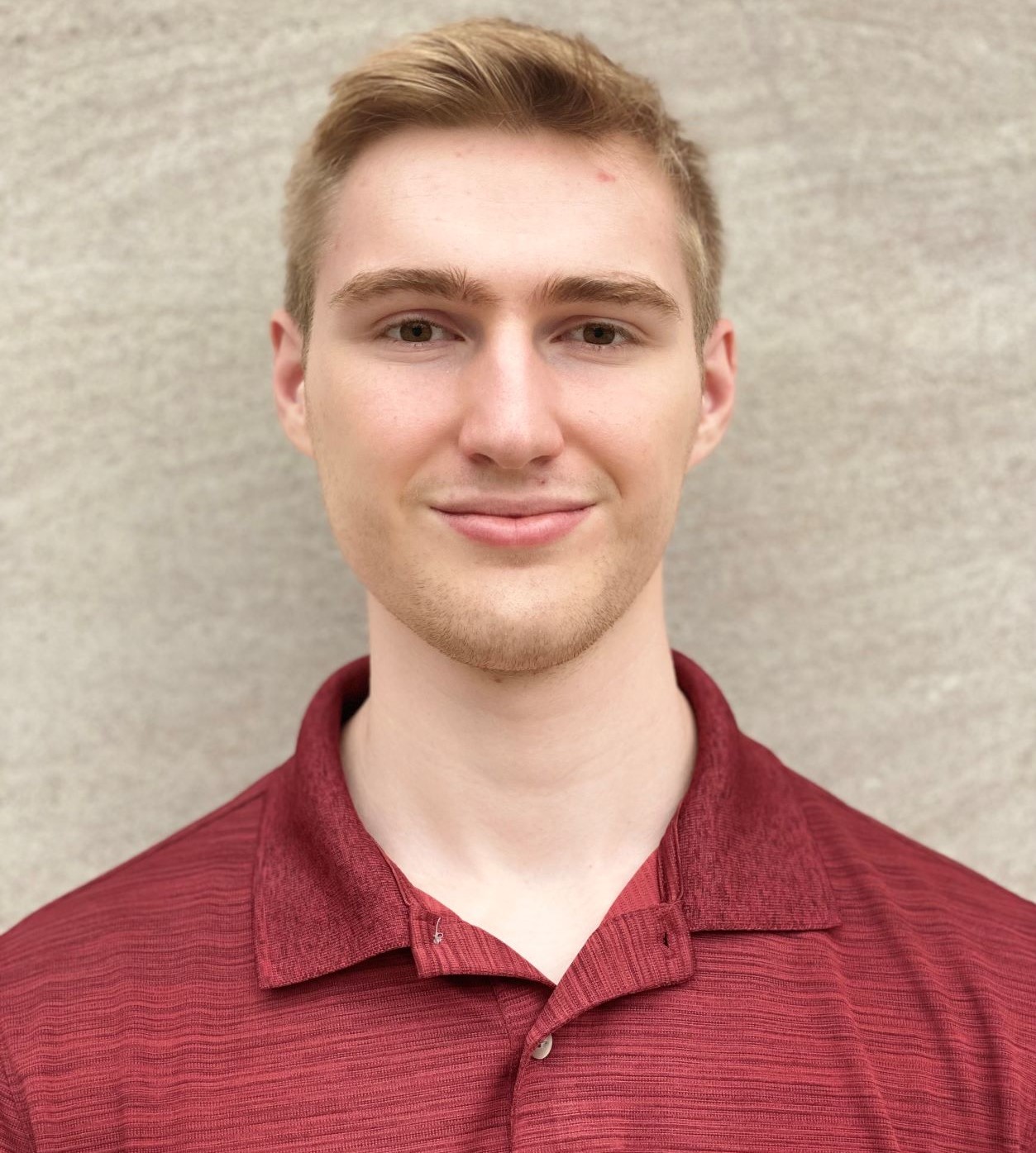}}]{\textbf{Daniel C. Stumpp}} 
received the Bachelor of Science degree in electrical engineering from the University of Pittsburgh in 2020. He is currently pursuing M.S. and Ph.D. degrees in electrical and computer engineering at the University of Pittsburgh. He is a member of the NSF Center for Space, High-Performance, and Resilient Computing (SHREC), where he performs research under the direction of Dr. Alan George. His research interests include high-performance FPGA based accelerator architectures for embedded platforms and novel processing of asynchronous neuromorphic sensor data. 
\end{IEEEbiography}

\begin{IEEEbiography}[{\includegraphics[width=1in,height=1.25in,clip,keepaspectratio]{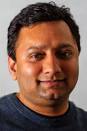}}]{\textbf{Himanshu Akolkar}}
received the M.Tech. degree in electrical engineering from IIT Kanpur, Kanpur, India, and the Ph.D. degree in robotics from IIT Genoa, Genoa, Italy, after which he had a postdoctoral position at Universite Pierre et Marie Curie. He is currently a Postdoctoral Associate with the University of Pittsburgh. His primary research interest includes the neural basis of sensory and motor control to develop an intelligent machine.
\end{IEEEbiography}

\begin{IEEEbiography}[{\includegraphics[width=1in,height=1.25in,clip,keepaspectratio]{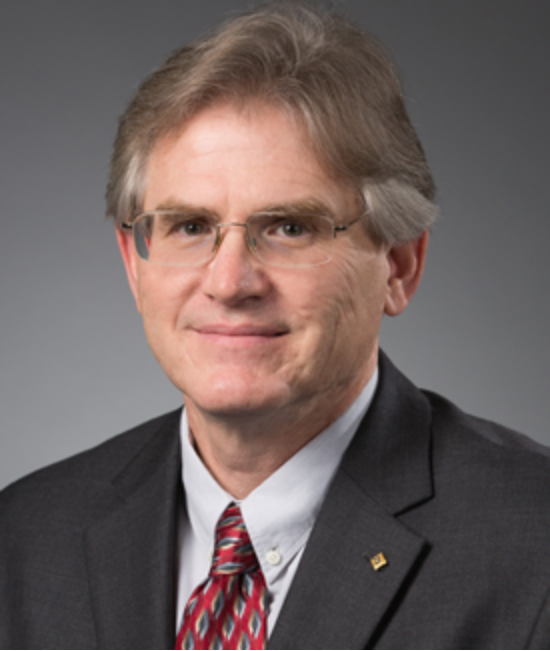}}]{\textbf{Alan D. George}}
(Fellow, IEEE) is currently the Department Chair, the Robert Horonjeff Mickle Endowed Chair, and a Professor of electrical and computer engineering (ECE) with the University of Pittsburgh. He is the Founder and the Director of the NSF Center for Space, High-Performance, and Resilient Computing (SHREC) headquartered at Pittsburgh. SHREC is an industry/university cooperative research center (I/UCRC) featuring some 30 academic, industry, and government partners and is considered by many as the leading research center in its field. His research interests include high-performance architectures, applications, networks, services, systems, and missions for reconfigurable, parallel, distributed, and dependable computing, from spacecraft to supercomputers. He is a Fellow of the IEEE for contributions in reconfigurable and high-performance computing.
\end{IEEEbiography}

\begin{IEEEbiography}[{\includegraphics[width=1in,height=1.25in,clip,keepaspectratio]{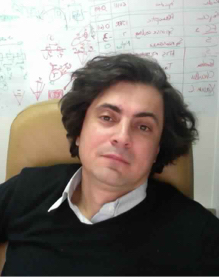}}]{\textbf{Ryad B. Benosman}}
received the M.Sc. and Ph.D. degrees in applied mathematics and robotics from University Pierre and Marie Curie, in 1994 and 1999, respectively. He is currently a Full Professor with the University of Pittsburgh/Carnegie Mellon/Sorbonne University. His work pioneered the field of event-based vision. He is also the Co-Founder of several neuromorphic related companies, including Prophesee—the world leader company in event-based cameras, Pixium Vision—a retina prosthetics company. He has authored more than 60 publications that are considered foundational to the field of event-based vision. He holds several patents in the area of event vision, robotics, and image sensing. In 2013, he was awarded with the national best French Scientific article by the publication LaRecherche for his work on neuromorphic retinas applied to retina prosthetics.
\end{IEEEbiography}

\EOD

\end{document}